\documentclass[10pt, conference, letterpaper]{ IEEEtran}

\IEEEoverridecommandlockouts
\usepackage[ruled,linesnumbered]{algorithm2e}
\usepackage{multirow} 
\usepackage{booktabs}
\usepackage{graphicx}
\usepackage{textcomp}
\usepackage{xcolor}
\usepackage{makecell}
\usepackage{amsmath}
\usepackage{amsthm,amssymb,amsfonts}
\usepackage{subcaption}
\usepackage{tikz}
\usepackage{listings}
\usepackage{siunitx} 
\usepackage[hidelinks]{hyperref}

\begin{document}

\begin{sloppypar}
\title{The Small-World Beneath LEO Satellite Coverage: Ground Hubs in Multi-Shell Constellations
%{\footnotesize \textsuperscript{*}Note: Sub-titles are not captured for https://ieeexplore.ieee.org  and
%should not be used}
%\thanks{Identify applicable funding agency here. If none, delete this.}
}

\author{Hailong Su,
	Jinshu Su,~\IEEEmembership{Senior~Member,~IEEE, 
        Yusheng Xia,
        and~Haibin Li
	  }% <-this % stops a space
\thanks{H. Su is with School of Electronic Information and Electrical Engineering, Shanghai Jiaotong University, Shanghai, China (e-mail: hailongsu@sjtu.edu.cn).}
\thanks{J. Su, Y. Xia (corresponding author) are with Chinese Academy of Military Science, Beijing, China (xys@nudt.edu.cn).}
 \thanks{J. Su is with College of Computer Science and Technology, National University of Defense Technology (NUDT), Hunan, China (e-mail: sjs@nudt.edu.cn).}
  \thanks{H. Li is with College of Information and Communication, NUDT, Hubei, China (e-mail: lihb15@tsinghua.org.cn).}

}

%\author{\IEEEauthorblockN{1\textsuperscript{st} Given Name Surname}
%\IEEEauthorblockA{\textit{dept. name of organization (of Aff.)} \\
%\textit{name of organization (of Aff.)}\\
%City, Country \\
%email address or ORCID}
%\and
%\IEEEauthorblockN{2\textsuperscript{nd} Given Name Surname}
%\IEEEauthorblockA{\textit{dept. name of organization (of Aff.)} \\
%\textit{name of organization (of Aff.)}\\
%City, Country \\
%email address or ORCID}
%\and
%\IEEEauthorblockN{3\textsuperscript{rd} Given Name Surname}
%\IEEEauthorblockA{\textit{dept. name of organization (of Aff.)} \\
%\textit{name of organization (of Aff.)}\\
%City, Country \\
%email address or ORCID}
%\and
%\IEEEauthorblockN{4\textsuperscript{th} Given Name Surname}
%\IEEEauthorblockA{\textit{dept. name of organization (of Aff.)} \\
%\textit{name of organization (of Aff.)}\\
%City, Country \\
%email address or ORCID}
%\and
%\IEEEauthorblockN{5\textsuperscript{th} Given Name Surname}
%\IEEEauthorblockA{\textit{dept. name of organization (of Aff.)} \\
%\textit{name of organization (of Aff.)}\\
%City, Country \\
%email address or ORCID}
%\and
%\IEEEauthorblockN{6\textsuperscript{th} Given Name Surname}
%\IEEEauthorblockA{\textit{dept. name of organization (of Aff.)} \\
%\textit{name of organization (of Aff.)}\\
%City, Country \\
%email address or ORCID}
%}

\maketitle

\begin{abstract}
In recent years, the emergence of large-scale Low-Earth-Orbit (LEO) satellite constellations has introduced unprecedented opportunities for global connectivity. However, routing efficiency and inter-shell communication remain key challenges in multi-shell architectures. This paper investigates the structural properties and network dynamics of a representative six-shell mega-constellation composed of 10,956 satellites and 198 gateway stations (GSs). Leveraging tools from complex network analysis, we identify several critical findings: (1) the constellation exhibits strong small-world characteristics, enabling efficient routing despite large network diameters; (2) GS relays play a pivotal role in enhancing inter-shell connectivity by bridging otherwise disconnected components; (3) feeder links significantly reduce average path length, making long-haul communication more feasible; (4) betweenness analysis reveals load imbalances among GSs, indicating the need for traffic-aware management strategies; (5) the architecture offers excellent spatial coverage and resilience, maintaining connectivity and low routing costs even under GS failures. These insights not only explain the design rationale behind current mega-constellations like SpaceX Starlink, but also provide valuable guidance for the evolution of future satellite network infrastructures.
\end{abstract}


\section{Introduction}
\label{sec:intro}
With the acceleration of globalization and the rapid advancement of information technology, globally interconnected communication has become a cornerstone of modern society. However, geographical constraints and the complexity of natural environments continue to pose significant challenges, including limited network coverage and difficulties in deploying wireless communication infrastructure in remote and underserved regions \cite{1}. To overcome these limitations, the focus of sixth-generation (6G) wireless network research has increasingly shifted toward non-terrestrial networks (NTNs), with the goal of enabling truly seamless global connectivity \cite{2}. 

Since the introduction of the Low-Earth-Orbit (LEO) satellite constellation network concept in the 1990s, several companies—such as Teledesic, Celestri, and Skybridge—have pursued its development. Although many of these early initiatives ended in bankruptcy or project abandonment, they shared a common vision: to deliver low-latency, high-bandwidth, and wide-coverage network services \cite{3}. Today, fueled by advances in space technology, Starlink—developed by SpaceX—is widely regarded as the most promising and large-scale implementation of this vision \cite{4}. In addition, project Kuiper \cite{project_kuiper}, led by Amazon, began full-scale deployment in April 2025 and is widely regarded as a strong competitor to Starlink. Leveraging the global infrastructure of Amazon Web Services.

\begin{table*}[htbp]
\centering
\caption{Orbital Parameters of Starlink Gen 1 and Gen 2 Constellations}
\small
\begin{tabular}{@{}llccccc@{}}
\toprule
\textbf{Generation} & \textbf{Shell} & \textbf{Altitude (km)} & \textbf{Inclination (°)} & \textbf{Orbital Planes} & \textbf{Satellites per Plane} & \textbf{Total Satellites} \\
\midrule
Gen 1 
      & S1 & 550 & 53.0  & 72 & 22 & 1,584 \\
      & S2 & 540 & 53.2  & 72 & 22 & 1,584 \\
      & S3 & 570 & 70.0  & 36 & 20 & 720 \\
      & S4 &560 & 97.6  & 6  & 58 & 348 \\

\midrule
Gen 2 & S5 & 530        & 43.0  & 56 & 60 & 3,360 \\
      & S6 & 535        & 33.0  & 56 & 60 & 3,360 \\
\bottomrule
\end{tabular}
\label{tab:starlink_generations}
\end{table*}
\begin{figure*}[ht]
    \centering
    \begin{subfigure}{0.15\textwidth}
        \includegraphics[width=\linewidth]{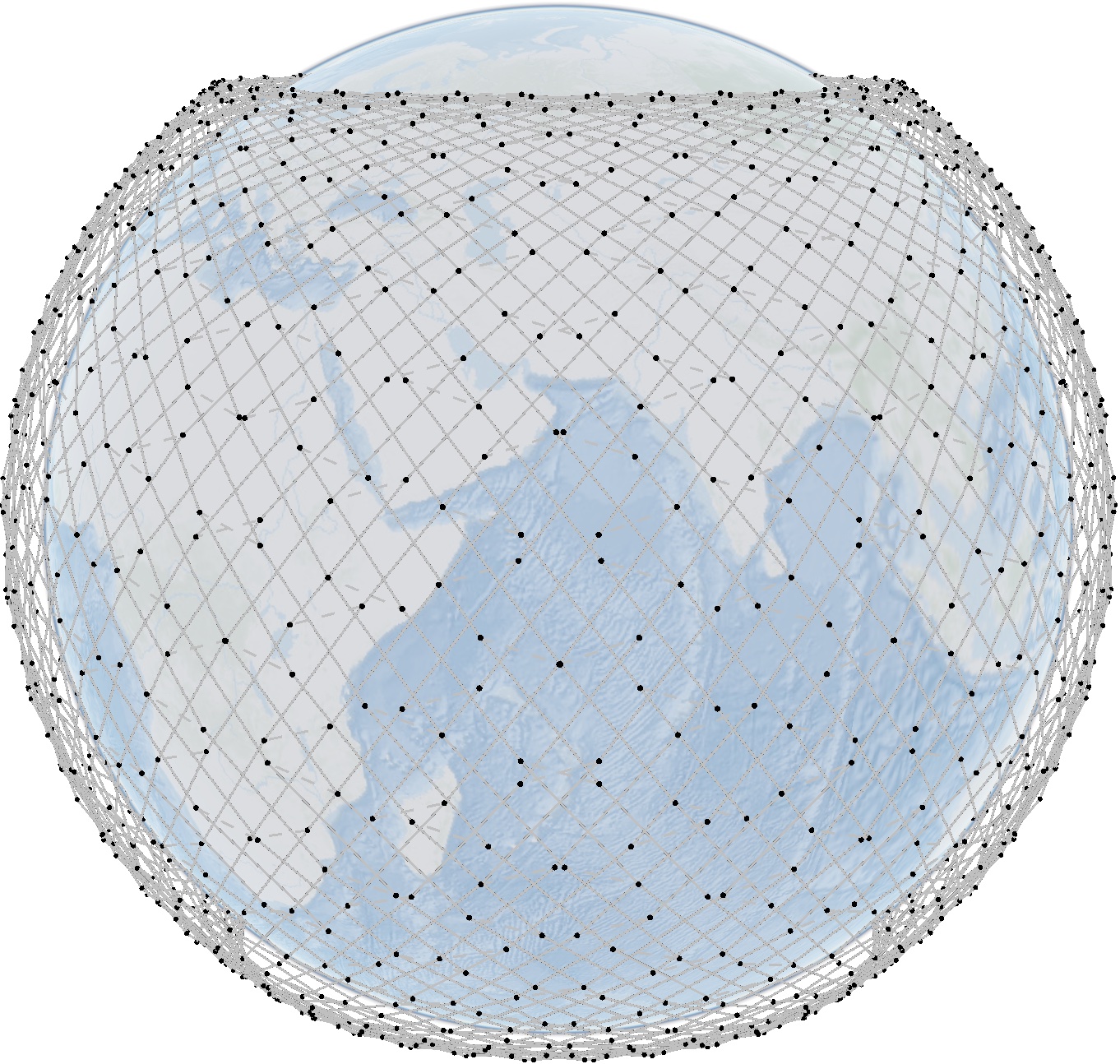}
        \caption{StarLink S1}
    \end{subfigure}\hspace{0pt}
    \begin{subfigure}{0.15\textwidth}
        \includegraphics[width=\linewidth]{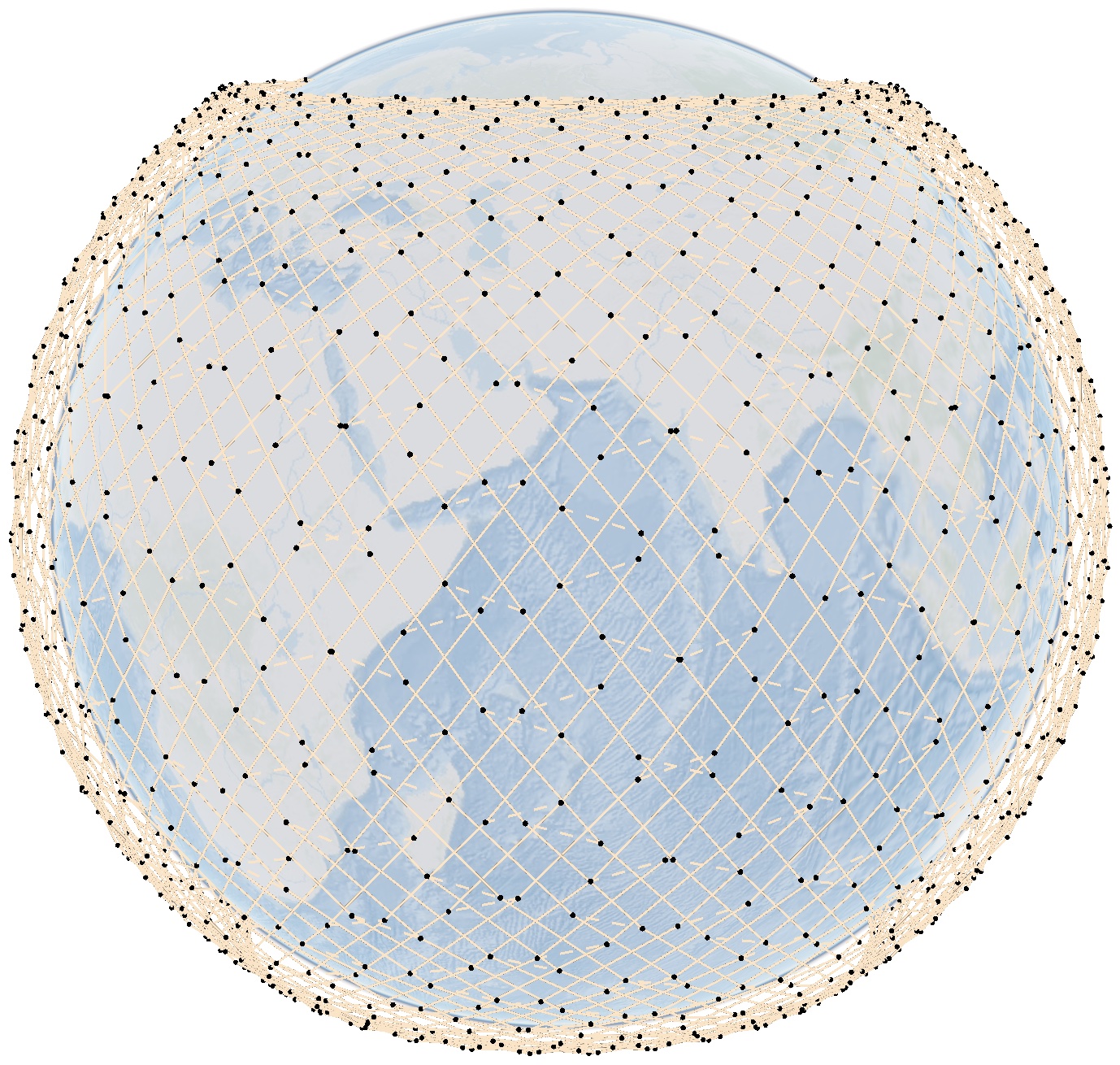}
        \caption{StarLink S2}
    \end{subfigure}\hspace{0pt}
    \begin{subfigure}{0.15\textwidth}
        \includegraphics[width=\linewidth]{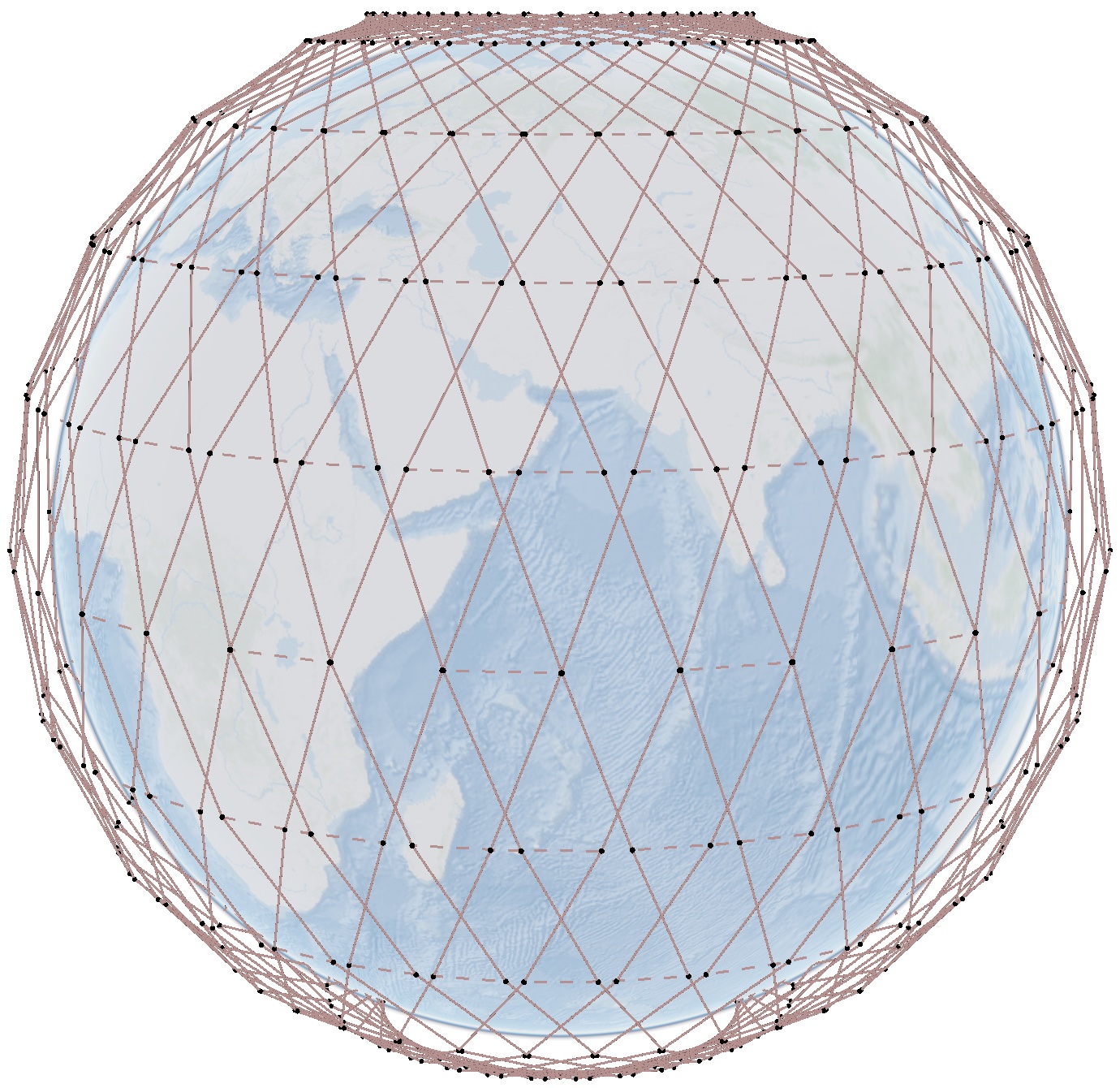}
        \caption{StarLink S3}
    \end{subfigure}\hspace{0pt}
    \begin{subfigure}{0.145\textwidth}
        \includegraphics[width=\linewidth]{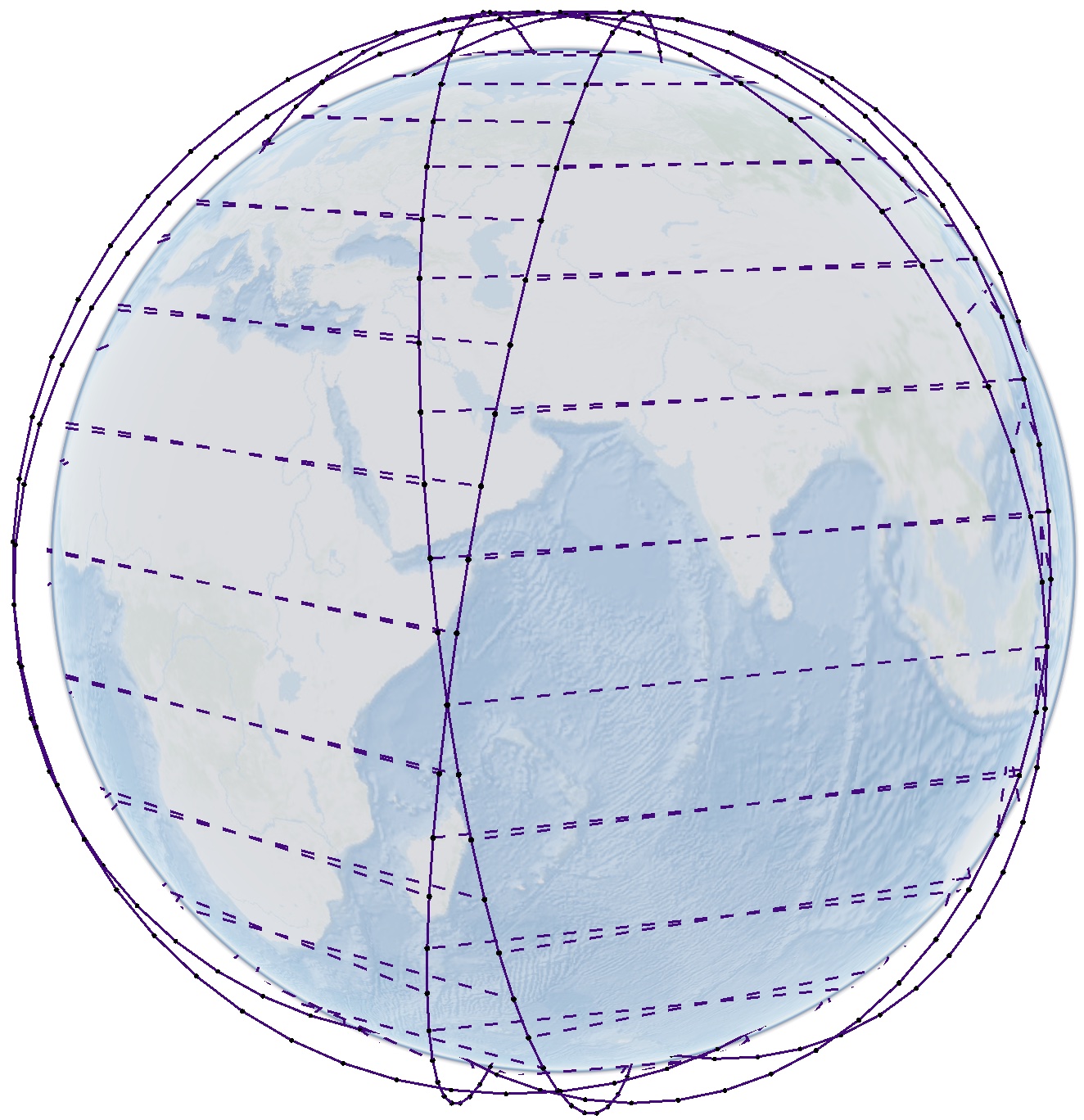}
        \caption{StarLink S4}
    \end{subfigure}\hspace{0pt}
    \begin{subfigure}{0.16\textwidth}
        \includegraphics[width=\linewidth]{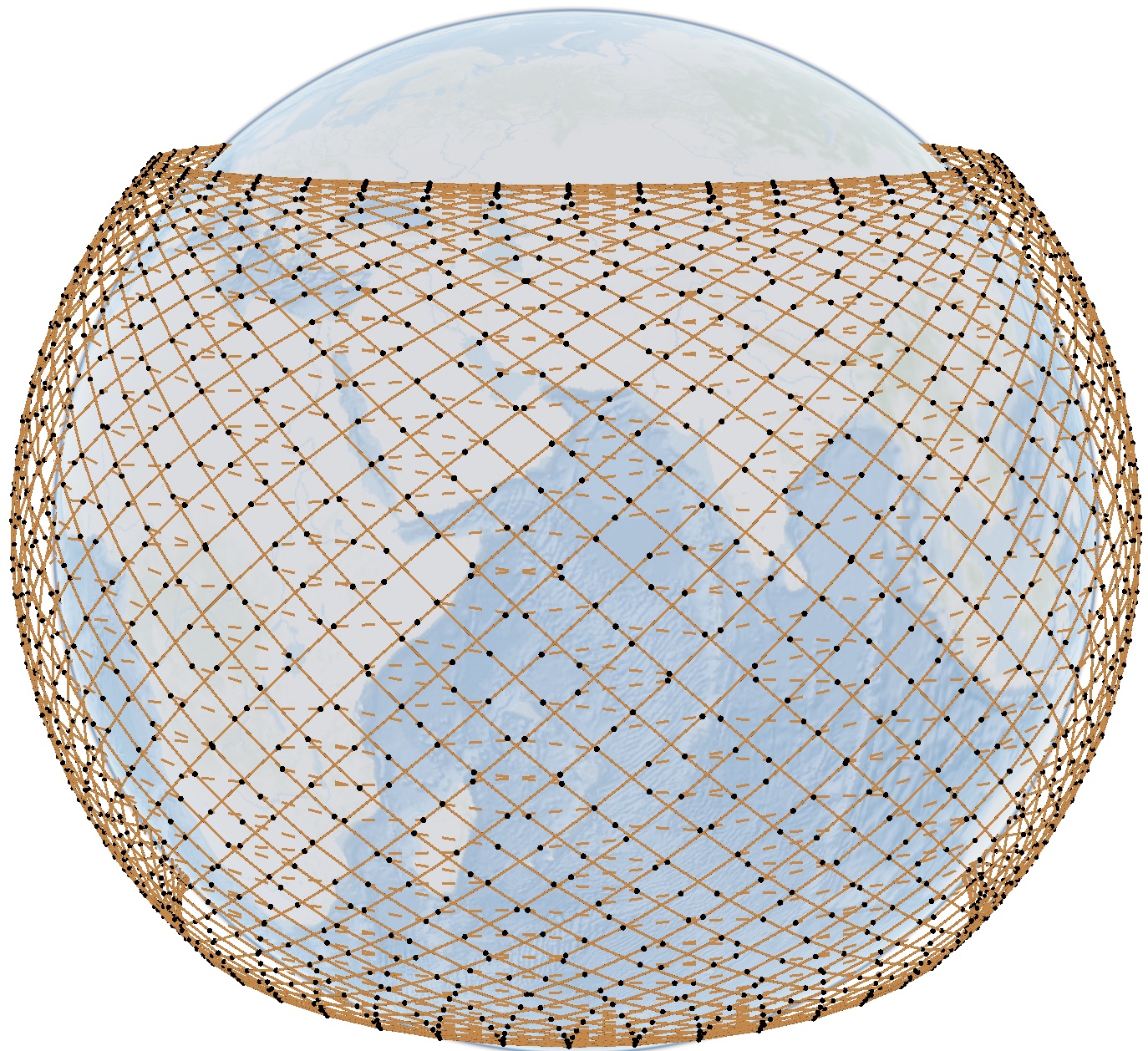}
        \caption{StarLink S5}
    \end{subfigure}\hspace{0pt}
    \begin{subfigure}{0.16\textwidth}
        \includegraphics[width=\linewidth]{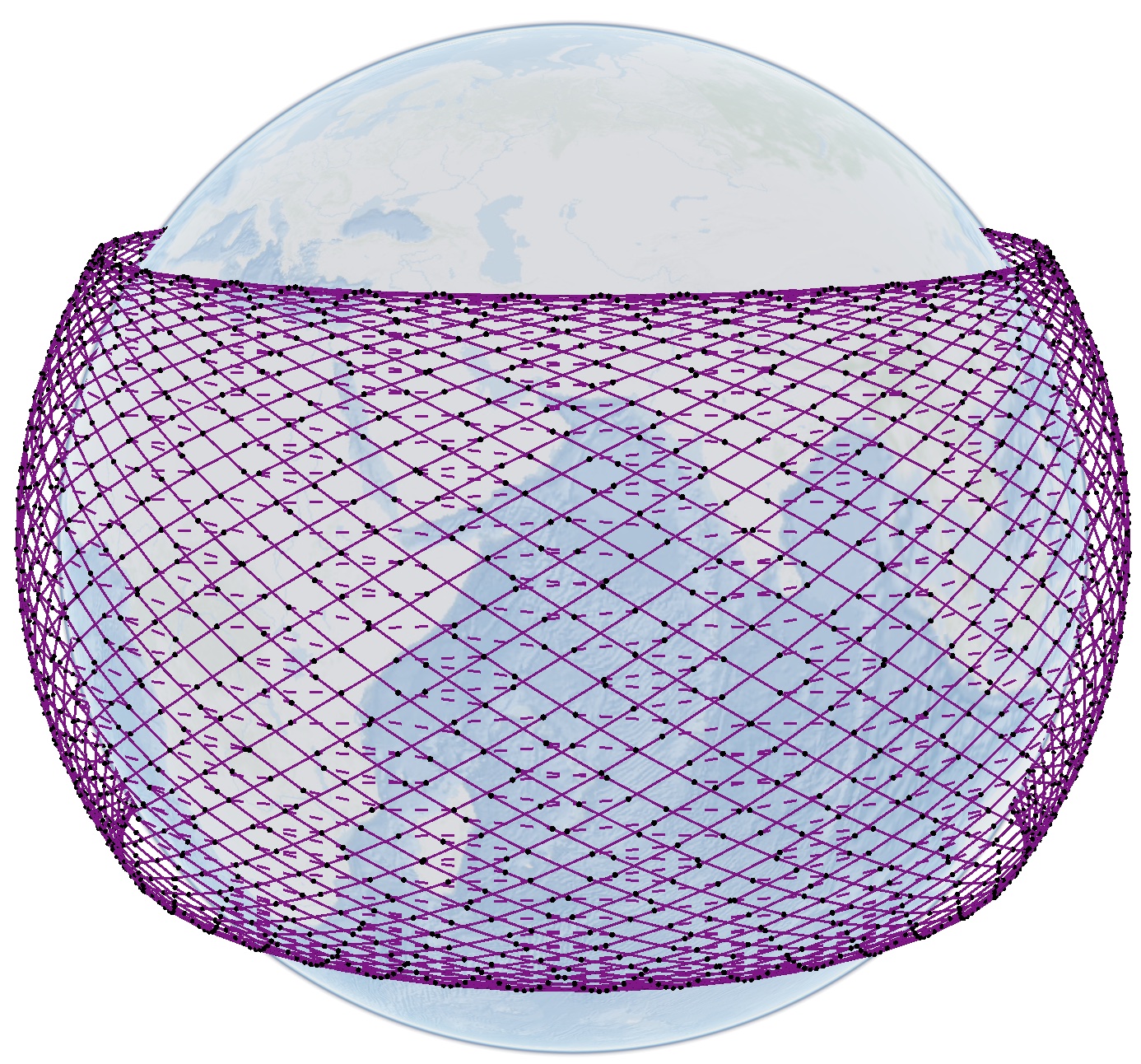}
        \caption{StarLink S6}
    \end{subfigure}
    \caption{
        Schematic diagram of the StarLink LEO satellite constellation plan. 
	The network diameters are 72, 72, 36, 32, 58 and 58, and the average path lengths are 36.58, 36.58, 18.94, 16.09, 33.36 and 33.36, respectively.
    }
    \label{Cons_Viz}
\end{figure*}

Starlink initially proposed a mega-constellation comprising 42,000 satellites. However, this ambitious plan has not yet been fully approved by the Federal Communications Commission (FCC). The first-generation (Gen 1) deployment, including versions v1.0 and v1.5, consists of 4,236 satellites distributed across orbital shells with inclinations of 53°, 53.2°, 70°, and 97.6°, and altitudes ranging from 540 km to 570 km \cite{10}. This phase has been largely completed. The configuration of the Gen 1 constellation is shown in Table~\ref{tab:starlink_generations}.

The Starlink communication model operates as follows: a ground user terminal (UT) connects to a satellite, which relays the signal to the nearest gateway station or gateway station (GS) that is part of a terrestrial gateway network. To receive service, users must be located within 500 miles of a GS \cite{betz2021starlink}.

The introduction of inter-satellite links (ISLs) reduces dependence on ground infrastructure, enabling internet access in regions without GS coverage. According to the 2024 Starlink progress report \cite{starlink2024}, each satellite is equipped with three sets of ISL devices. This configuration, aimed at controlling costs, contrasts with the commonly assumed grid-based ISL topology widely discussed in academic literature~\cite{grid}. After all, SpaceX must remain economically viable to sustain the continued expansion of its infrastructure footprint.

According to \cite{Starlink_Scheduling}, Starlink employs a global scheduler that operates every 15 seconds to assign serving satellites to UTs. This scheduler takes into account various factors such as elevation angle, azimuth, sunlight exposure, and satellite lifetime.

It remains unclear whether the scheduler distinguishes between shells when assigning satellites to UTs. However, it is evident that maintaining communication quality in high-latitude regions depends heavily on high-inclination sub-constellations like shell 3 and shell 4. In fact, most low-inclination shells are not even visible from these regions.

This raises an \textbf{important question:} if two ground nodes (either UTs or GSs) are connected to satellites in different shells, is cross-shell communication required to establish end-to-end connectivity? Or, alternatively, does the resource scheduler enforce a constraint that both the source terminal and the destination GS must connect to satellites within the same shell to ensure basic network reachability?

If relaying is disallowed and the system mandates that both the ingress and egress satellites reside within the same shell, user traffic must traverse only via ISLs and be handed off to the gateway through a downlink at the final hop. Under such a constraint, routing intercontinental traffic can incur prohibitively high costs—particularly for dense sub-constellations such as S1, S2, S5, and S6.

% LEO satellite constellations claim to provide communication between any two points on the globe. A pertinent question arises: how can such a system facilitate real-time communication between latitudinal distant locations, such as Singapore and Reykjavik? Furthermore, can researchers stationed at the South Pole utilize StarLink to access the World Wide Web? If two ground-based network nodes are connected to different shells, several methods can be considered for their communication: 
If cross-shell communication is required, several relay mechanisms can be considered:
\textbf{Geostationary and Medium Earth Orbit (GEO/MEO) Satellites:} Communication via GEO satellites involves relaying signals through satellites positioned approximately 35,786 kilometers above the Earth’s equator, while MEO satellites operate at altitudes between 2,000 and 35,786 kilometers. Although both options offer wide coverage, they introduce significant round-trip time (RTT)—approximately 240 milliseconds for GEO, and variable but still substantial for MEO. 

\textbf{Inter-Shell ISLs:} In theory, ISLs using laser communication could enable direct communication between different orbital shells. However, the acquisition, tracking, and pointing (ATP) mechanisms required for laser links \cite{7} are severely challenged by the high relative velocities between satellites in different shells, making it difficult to establish and maintain stable links \cite{8}. In addition, this approach has been explicitly abandoned, as stated in documents submitted by Starlink to the FCC \cite{SpaceX2018FCC}.

\textbf{Ground Relays:} A more practical approach involves using ground relay stations to facilitate communication across shells. In this scenario, data from a terminal is first transmitted to a satellite, then downlinked to a relay GS (potentially involving multiple terrestrial relays along the shortest path), uplinked to a satellite in the target shell, and finally delivered to the destination gateway station. This method leverages existing terrestrial infrastructure, avoids the complexity associated with high-velocity inter-shell ISLs, and improves the routing efficiency.

All available evidence suggests that utilizing GS relays is the most effective solution for reducing routing costs \cite{starlinkinfo2025, starlinkinfooffical2025}. This paper \textbf{aims to uncover the underlying rationale from the perspective of complex network analysis}. In addition, we propose several insightful observations and forward-looking considerations that carry important implications for the future design and evolution of satellite constellation networks. 

 In our dynamic study of multi-shell constellation topologies (10,956 satellites in 6 shells) enhanced by 198 GSs, we make the following key findings:

\begin{itemize}
    \item \textbf{Small-world property:} By analyzing the degree distribution and clustering coefficient, we firstly observe that mega-constellations exhibit small-world characteristics, which are indicative of efficient global connectivity despite large network diameters.
    
    \item \textbf{Connectivity enhancement:} The introduction of GS relays enables feeder links to act as bridges, interconnecting satellites from different shells and effectively forming a fully connected component across the constellation.
    
    \item \textbf{Routing efficiency:} Although mega-constellations possess large network diameters, the average path length remains remarkably low—close to that of the sparsest single-shell constellations—revealing substantial potential for high-efficiency routing.

    \item \textbf{Load balancing challenges:} Through betweenness analysis, we find that GSs serve as critical hubs within the network. However, a large disparity in load exists among different GSs, posing significant challenges for network management and traffic engineering.
    
    \item \textbf{Global coverage:} Multi-shell constellations significantly improve signal quality and service availability for UTs worldwide by offering better spatial coverage and a larger pool of candidate satellites for access.

    \item \textbf{Robustness:} Even after removing a portion of GSs, the network maintains its average path length and connectivity, highlighting the inherent robustness of the mega-constellation architecture.
\end{itemize}

Moreover, the above conclusion holds across varying constellation shells, ISL configurations, and satellite handover strategies. We advocate that future LEO satellite constellations should \textbf{treat Inter-Satellite Links (ISLs) and Ground-Satellite Links (GSLs) as a unified routing resource}, enabling coordinated, cross-layer, end-to-end shortcut communication. Relying solely on GSLs in a bent-pipe architecture, or depending exclusively on ISLs until the final downlink to a ground station, both fail to achieve optimal routing efficiency.

\section{Background}
\label{sec:related}
\subsection{Crucial Role of Gateway Stations}
Gateway stations are essential in LEO constellations, serving as primary communication hubs between satellites and terrestrial networks. They handle data uplinks and downlinks, perform network management tasks, and facilitate satellite handovers \cite{25, 34}. 

LEO constellations use two primary communication forms: laser communication for ISLs and radio frequency (RF) for Ground-Satellite Links (GSLs) or feeder links (FLs). Laser communication offers significantly higher bandwidth, making it ideal for ISLs. RF communication for FLs faces challenges like signal attenuation due to atmospheric conditions. The downlink capacity of FLs ranges from 17 to 23 Gbps \cite{19}, while laser ISLs could exceed 100 Gbps \cite{20}. Although FLs offer lower bandwidth compared to ISLs, their significance \textbf{should not be underestimated}. 

Previous studies have focused on optimizing GS deployment using various algorithms. One study \cite{22} used improved genetic and simulated annealing algorithms but treated GSs as data endpoints rather than relays. Another \cite{21} proposed an iterative deployment algorithm based on marginal revenue maximization but only considered the bent-pipe model without ISLs. Additionally, a study \cite{24} employed evolutionary algorithms to optimize the number and placement of GSs, focusing on single-shell constellations. However, with the extensive existing GS infrastructure like Starlink, the focus shifts to scheduling rather than deployment optimization.
\subsection{Synergy between ISLs and FLs}

As highlighted in \cite{18}, omitting ISLs—despite dense GS deployments—leads to significantly higher latency variability, reduced resilience to weather disruptions, and a substantial drop in overall network throughput.

If the constellation relies heavily on FLs and uses multiple ground relays, it can create bottlenecks, limiting overall network performance. Thus, leveraging GSs as supplementary routes for traffic is more practical in the context of ISL deployment, despite increasing network complexity.

As literature \cite{28} shows, in a single shell constellation, gateway stations enhance routing efficiency by reducing the number of hops required for data transmission, acting as relay points that bypass multiple ISL hops. However, the characteristic of the small-world nature behind this has not been revealed.
 
In multi-shell constellations, maintaining network connectivity is a primary challenge due to the dynamic topology and varying altitudes of satellite shells. GSs are critical hubs that bridge gaps between satellite shells. 

\subsection{The Method of Complex Network Analysis}
Complex network analysis provides valuable tools for understanding the structural properties and performance characteristics of multi-shell LEO satellite constellations integrated with GSs. The foundational work by Watts and Strogatz \cite{watts1998collective} introduced the concept of small-world networks, highlighting how local clustering and short global path lengths can coexist—an idea highly relevant to satellite constellations.

Subsequently, Albert et al.\ \cite{26} initiated systematic investigations into the resilience of various types of complex networks, laying the groundwork for robustness studies in networked systems. In LEO constellations, the uniform spatial distribution of satellites leads to structural homogeneity, which limits the analytical insights that complex network theory can provide. However, when GSs are introduced to interconnect multiple homogeneous satellites, the resulting network exhibits significantly altered topological and functional characteristics.

Despite these compelling structural shifts, \textbf{research on complex network analysis applied to LEO satellite constellations remains sparse}. P. Gerald et al.\ \cite{27}, using percolation theory, studied robustness optimization in networks subject to random failures and showed that increased heterogeneity enhances resilience. This finding is particularly relevant to modern LEO satellite networks, where differences in GS connectivity introduce considerable node-level heterogeneity.
\section{Methodology}
\label{sec:meth}
\begin{figure}[ht]
\centering
\includegraphics[width=0.45\textwidth ]{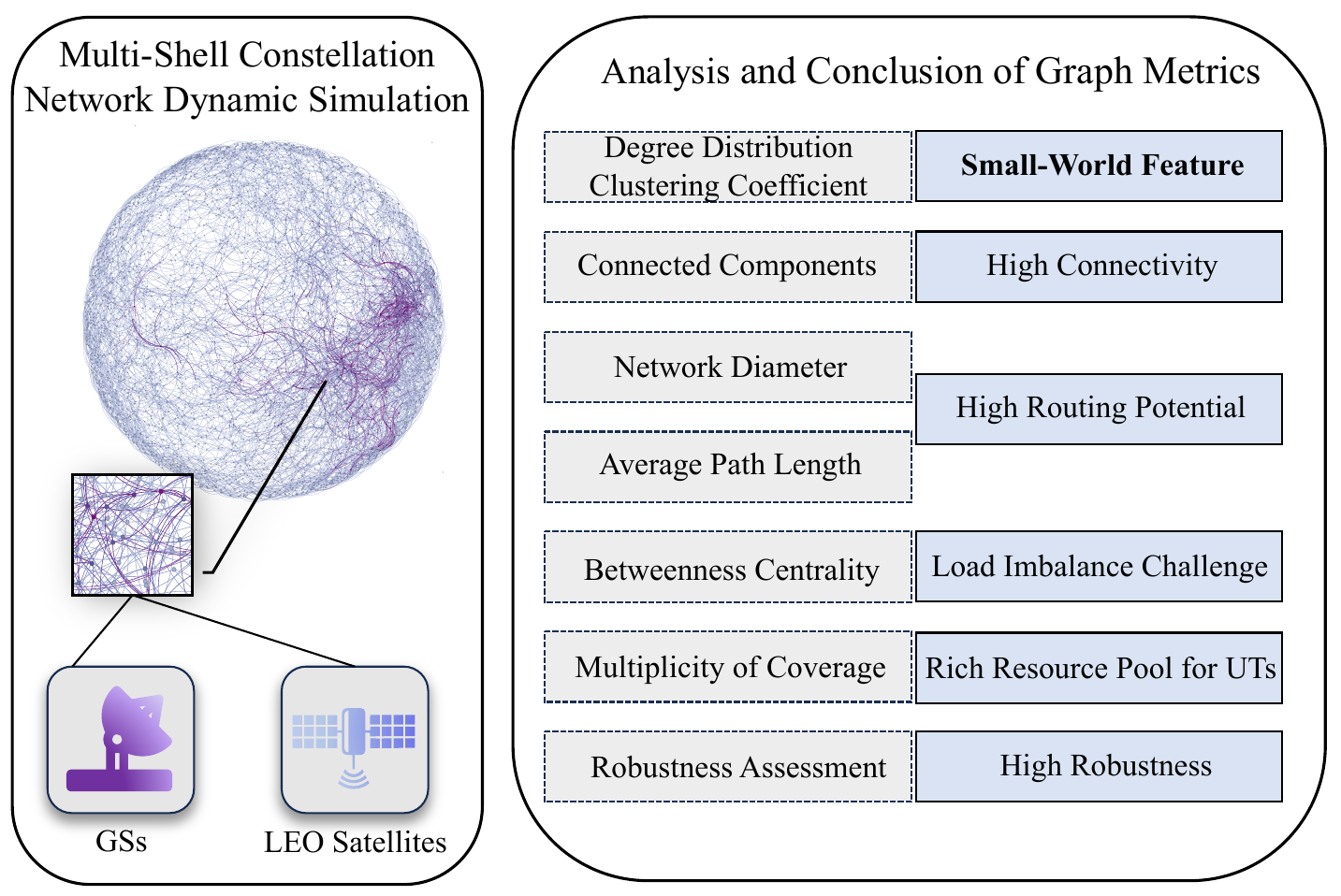}
\caption{Methodology of this study.}
\label{fig gephi}
\end{figure}

As Figure~\ref{fig gephi} shows, this study conduct a complex network analysis of a multi-shell constellation network with the integration of GS. 
To systematically evaluate the structural and functional properties of the integrated satellite-ground network, we adopt a complex network analysis approach based on key topological and geometric metrics. The simulation environment models a realistic multi-shell constellation comprising 10,956 LEO satellites and 198 gateway stations, resulting in a total of 11,154 nodes. The edges represent either ISLs or FLs, depending on elevation angle constraints and visibility rules.

We first analyze the degree distribution to assess heterogeneity, revealing structural variation due to shell and GS placement. Despite a modest average clustering coefficient, many weak or quasi-triangle patterns suggest latent modularity and small-world tendencies.

Connectivity analysis confirms that GS-enhanced topologies form a globally reachable network. The network diameter and average path length are then measured to assess the efficiency of end-to-end routing under minimal hop-count assumptions. Despite the vast scale of the network, we observe that average path lengths remain low—suggesting strong routing potential.

Betweenness is used to identify potential bottlenecks and critical relay nodes, especially among satellites directly connected to GSs. Our results reveal a load imbalance pattern, where certain GS-linked satellites bear disproportionately high transit loads—posing potential challenges for traffic engineering and resource management.

The coverage multiplicity is used to quantify the spatial coverage. Our results show that the multi-shell LEO constellation provides a large number of simultaneous satellite access options for users at any latitude across the globe.

Lastly, robustness tests under random GS failures show minimal impact on connectivity and average path length, confirming the network’s fault tolerance.

\subsection{System Modeling}
In late 2022, the FCC authorized the deployment of 7,500 satellites as part of Starlink’s second-generation (Gen 2) system, which has since entered the deployment phase \cite{satellitetoday_starlink_gen2_2022}. According to official filings, as of February 27, 2025, a total of 6,751 active Starlink satellites were operational in orbit \cite{starlink_demisability_2025}, indicating the ongoing deployment of both V2 and V2 Mini satellites.

Although detailed configuration parameters for the Gen 2 constellation have not been fully disclosed, documents submitted by SpaceX in October 2024 indicate that the approved 7,500 satellites are planned to be distributed across three orbital shells with altitudes of 475 km, 480 km, and 485 km, and inclinations of 53°, 43°, and either 28° or 32°, respectively \cite{spacex_gen2_modification_2024}. Each shell is authorized to contain up to 56 orbital planes. Notably, the FCC’s grant does not prescribe the specific allocation of satellites across the three shells, leaving the distribution at SpaceX’s discretion. Thus far, deployment has primarily focused on the 43° shell, likely due to its operational continuity with the existing 53° Gen 1 infrastructure \cite{speidel2024navigating}.

Without loss of generality, this study adopts a simulation configuration in which 3,360 satellites are placed in each of the 43° and 32° shells, resulting in a total of 10,956 satellites distributed across six orbital shells, as summarized in Table~\ref{tab:starlink_generations} and Figure~\ref{Cons_Viz}. This forward-looking configuration provides greater relevance for next-generation satellite network research.

Each satellite can establish connections with up to \( N_{\text{max}_{s,g}} \) GSs, while each GS can simultaneously maintain links with up to \( N_{\text{max}_{g,s}} \) LEO satellites. GSs adopt a maximum elevation angle strategy for satellite selection. If the elevation angle of a satellite falls below the predefined thresholds \( \theta_{g_{\text{thre}}} \) for GSs, the respective node will switch to an alternative satellite offering the highest available elevation angle to maintain connectivity.

In the context of complex network analysis, UTs are treated as unstable nodes due to their on-demand and spatially dynamic behavior, and are thus excluded from the structural evaluation of the network. Table~\ref{tab:symbols} summarizes the key notations used in this paper.
\begin{table}[h]
\caption{Notations and Descriptions in This Paper}
    \centering
    \begin{tabular}{c|c}
         \hline
        \textbf{Notation} & \textbf{Description} \\
         \hline
        $D$ & Network diameter \\
        \hline
        $APL$ & Average path length\\
        \hline
        $BC_{v}$ & Betweenness centrality of the node $v$\\
 	\hline
        $N_{max_{g,s}}$ & Maximum number of satellites a GS can access\\
         \hline
        $N_{max_{s,g}}$ & Maximum number of GSs a satellite can access\\
         \hline
        $\theta_{g_{thre}}$ & Elevation angle threshold of GSs \\
        \hline
    \end{tabular}
    \label{tab:symbols}
\end{table}

%\begin{figure}[ht]
%\centering
%\includegraphics[width=0.48\textwidth ]{system modeling.pdf}
%\caption{The frame of system modeling.}
%\label{fig2}
%\end{figure}

\begin{figure}[ht]
\centering
\includegraphics[width=0.42\textwidth ]{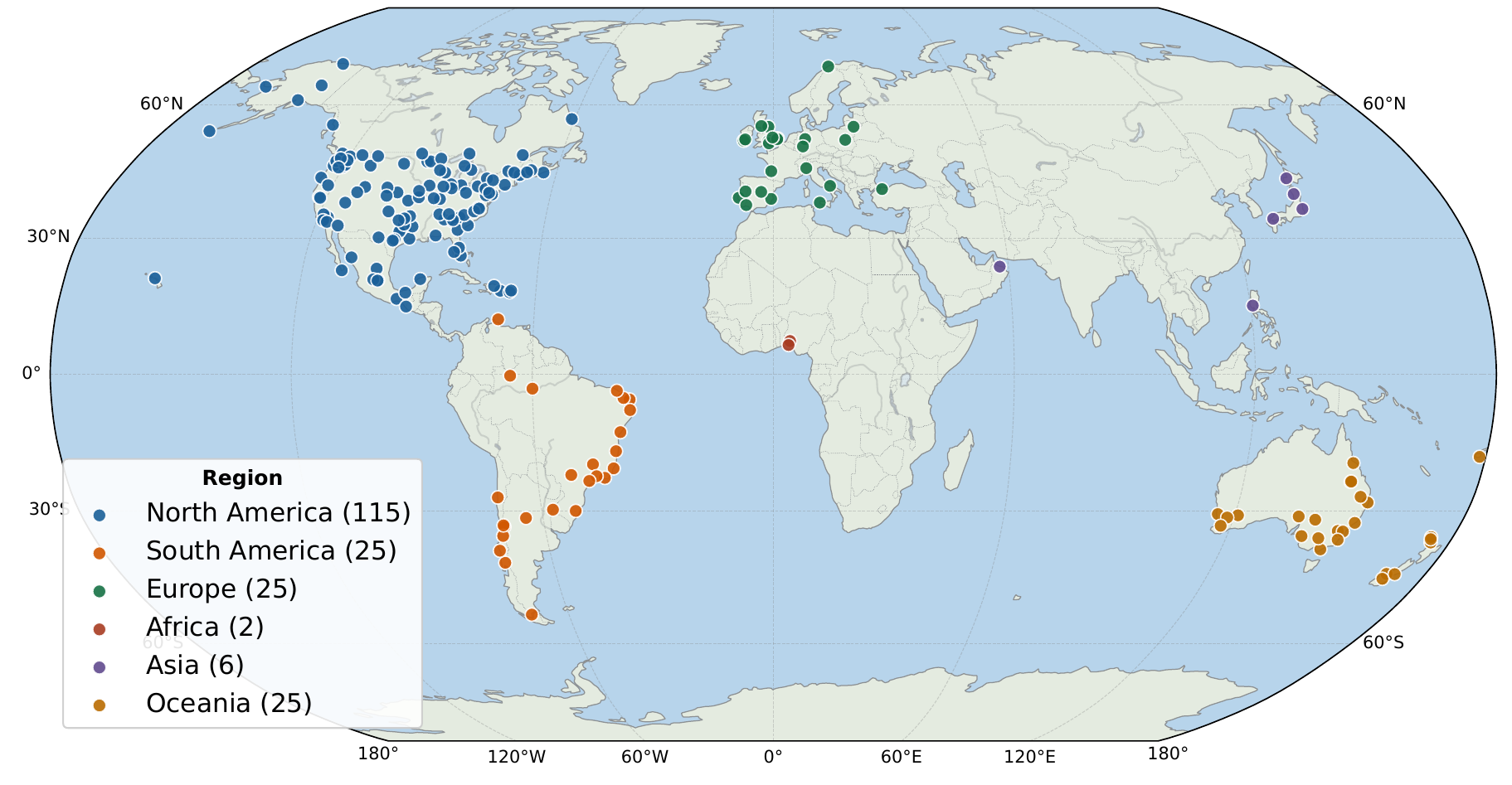}
\caption{Starlink Gateway Station Deployment Diagram.}
\label{fig3}
\end{figure}

A GS can typically maintain simultaneous connections with multiple satellites; however, the actual number is constrained by several factors, including line-of-sight visibility, radio frequency interference, antenna resources, beamforming capacity, scheduling strategies, and regulatory limitations. According to \cite{array}, each gateway is equipped with eight technically identical 1.5-meter parabolic antennas. In this study, we set the parameters \( N_{\text{max}_{g,s}} = 8 \), \( \theta_{g_{\text{thre}}} = 25^\circ \).

Currently, there is no official specification regarding the maximum number of GSs to which a satellite can simultaneously connect. This parameter is generally influenced by satellite hardware constraints and traffic engineering policies. Community-based link budget analyses—such as those conducted by Reddit users \cite{reddit_gateway_2021}—suggest that a single GS can deliver approximately 8–11 Gbps under realistic modulation and coding schemes. Given that a satellite’s downlink capacity may reach up to 20 Gbps, it is inferred that at least two GSs are required to fully utilize a satellite’s bandwidth. Based on this insight, we treat \( N_{\text{max}_{s,g}} \) as a tunable parameter to explore how feeder link capacity of satellites impacts overall network performance.

As shown in Figure~\ref{fig3}, these sites represent the 198 Starlink gateway stations that are either planned, under construction, or already operational (as of 2025 \cite{starlinkinsider2025}). For the purpose of this study, we make a forward-looking assumption that all listed stations are fully constructed and operational, and thus use them as the simulation baseline. The LEO satellite motion model known as simplified general perturbations 4 (SGP4) is capable of determining the precise position and velocity of satellites at any given time \cite{29}. A link between a satellite node and a ground node can only be established if the elevation angle and visibility requirements are met. All links are considered to be full-duplex, meaning that the network will form an undirected graph.

The simulation duration is set to 10,000 seconds, approximately twice the orbital period of S3—the shell with the highest altitude and, consequently, the longest orbital period in the constellation. The dynamic behavior of the constellation is simulated in ns-3 \cite{ns3_manual}, while graph metrics are measured using the of igraph based on C/C++ programming language \cite{CsardiNepusz2006}, ensuring scalability and computational efficiency for large-scale network simulations.

\subsection{Metrics and Explanation}
\label{sec:Metrics}

\textbf{Average Degree} in network theory represents the average number of connections each node maintains within a network. In a LEO satellite constellation network, it quantifies the typical number of links held by either a satellite or a GS. 

\textbf{Clustering Coefficient} in network theory measures the tendency of nodes to form tightly-knit groups, quantifying how likely a node's neighbors are to be connected with each other. In a LEO satellite constellation network, it captures the extent of local interconnectivity among satellites or GSs, revealing whether communication links tend to form localized clusters. 

\textbf{Connected Components} refer to subsets of nodes within a network where each node is reachable from any other node in the same subset~\cite{36}. The presence of multiple connected components indicates isolated nodes or subnetworks, which may lead to communication breakdowns. This study particularly focuses on whether satellites situated in different shells---otherwise unreachable to each other---can achieve connectivity through GSs.

\textbf{Network Diameter} is defined as the longest shortest path between any two nodes in the network~\cite{35}. Alongside, the \textbf{Average Path Length}, which is the average number of hops along the shortest paths between all pairs of nodes, is also considered. A smaller average path length indicates lower latency within the network.

\textbf{Betweenness} quantifies the extent to which a node lies on the shortest paths between other node pairs~\cite{37}. Nodes with high betweenness serve as critical intermediaries in the network; their failure can significantly disrupt connectivity and degrade data transmission efficiency. 

\textbf{Coverage Density} refers to the number of satellites simultaneously visible from a given ground location. A higher coverage density typically implies greater spatial diversity, enhanced link redundancy, and improved flexibility in satellite selection~\cite{38}. 

\section{Graph Analysis and Key Findings}
\subsection{Degree Distribution}
\begin{figure}[ht]
\centering
\includegraphics[width=0.45\textwidth ]{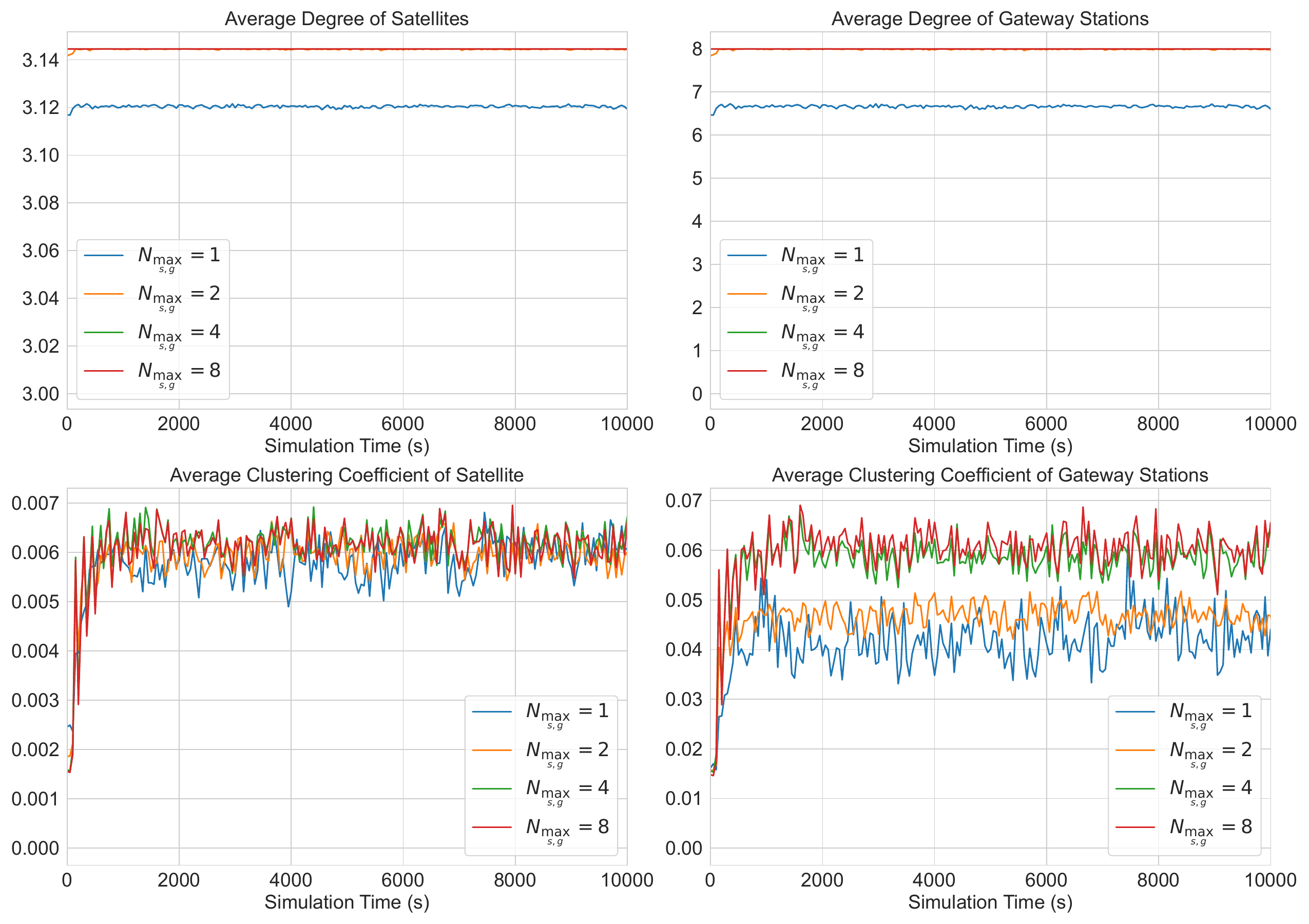}
\caption{The average degree and average clustering coefficient of satellites and gateway stations.}
\label{fig:degree}
\end{figure}

As shown in Figure~\ref{fig:degree} and consistent with expectations, the average degree of GSs is significantly higher than that of satellites, even when the satellite FL  capacity is increased to 8. The satellite population is approximately two orders of magnitude larger than the number of GSs, with over 10,000 satellites and only about 200 GSs available. As GSs are a limited resource, only a small subset of satellites can establish feeder links with them, while the majority are positioned over oceans, uninhabited regions, or other areas lacking ground infrastructure.

The number of satellites within the visibility range of a GS is limited, Due to the current concentration of GS deployments, when $N_{\max_{s,g}}$ is small, GSs may also compete for satellite resources, preventing some of them from fully utilizing all 8 parabolic antennas. However, when $N_{\max_{s,g}}$ is increased to 2, this contention is effectively eliminated.

While the constellation network includes high-degree nodes—such as GSs and satellites with feeder link connections—their presence is not sufficiently rare, as typically expected in scale-free networks \cite{barabasi1999emergence}. Instead, high-degree nodes are abundant and structurally redundant, with a relatively narrow disparity between the minimum and maximum degrees.

\begin{figure}[ht]
\centering
\includegraphics[width=0.45\textwidth ]{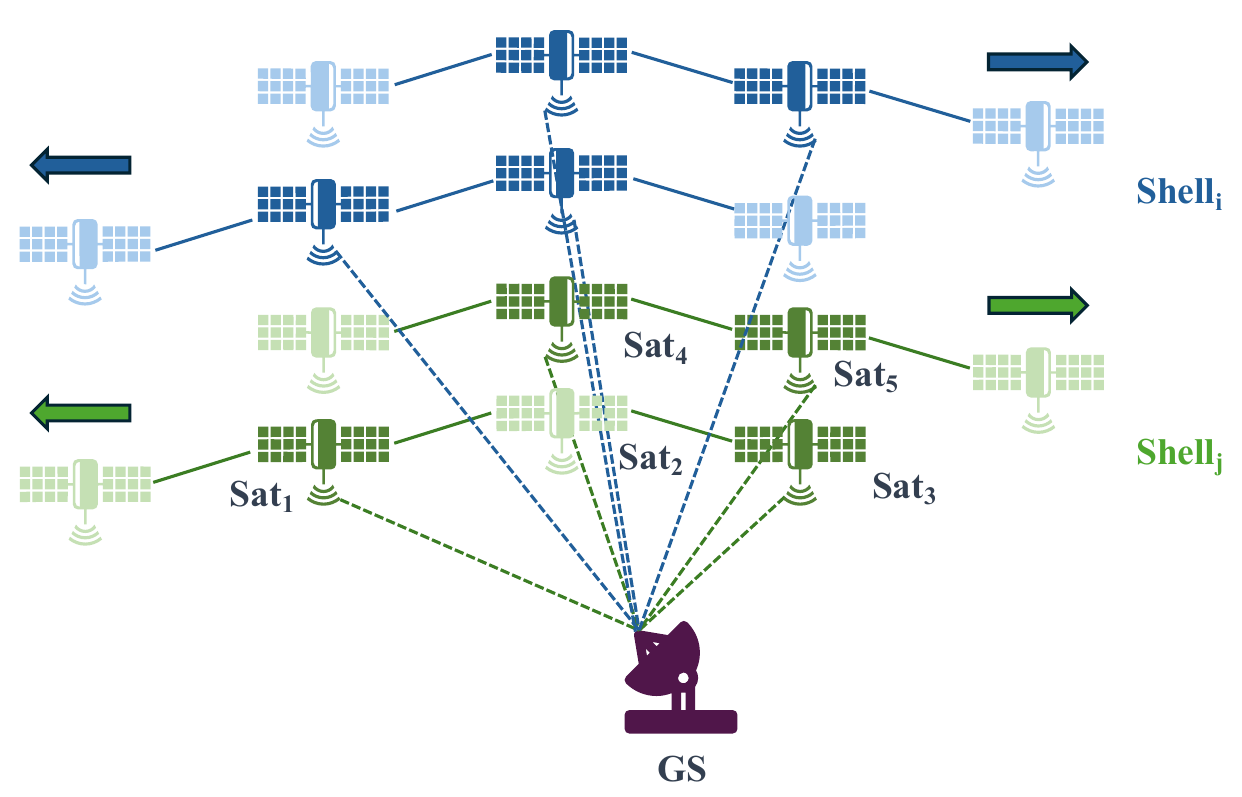}
\caption{Quasi-Triads and the Spatial-Topological mediation by gateway stations}
\label{fig4}
\end{figure}
\subsection{Clustering Coefficient Analysis and Nature of Small-World}
The core attributes of small-world networks are \textbf{short path lengths} and a \textbf{high clustering coefficient}. A high clustering coefficient signifies that there is a high probability for a node's neighbors to be interconnected.
Mathematically, the clustering coefficient $C$ for a node $i$ can be expressed as:
\begin{equation}
C_i = \frac{2E_i}{k_i(k_i - 1)},
\end{equation}
where $E_i$ represents the number of edges between the neighbors of node $i$, and $k_i$ is the degree of node $i$. For comparison, the clustering coefficient of an Erdős–Rényi random graph \cite{erdos1960evolution} with the same scale is approximately 0.001.

As shown in Figure~\ref{fig:degree}, the clustering coefficient of satellites is extremely low. This stems from the inherent structure of  ISLs, which—whether arranged in a grid or a 3-ISL configuration—intentionally avoids redundant connections. Although the clustering coefficient of GSs is roughly an order of magnitude higher than that of satellites, it still remains relatively low. Does this suggest that the constellation network lacks small-world properties? The answer is no.

As shown in Figure~\ref{fig4}, Sat4, Sat5, and the GS form a closed triangle, which satisfies the necessary condition for contributing to GS’s clustering coefficient calculation. Even though certain nodes (e.g., Sat1 and Sat3) do not form a strict triangle with the GS and thus do not contribute to its clustering coefficient, they are still able to communicate efficiently via a small number of intermediate satellites (e.g., Sat2) through high-bandwidth ISLs. Such quasi-triadic or near-triangular communication patterns are more prevalent in the constellation network. Consequently, although the clustering coefficient of GSs remains relatively low, the network can still exhibit small-world characteristics.

\subsection{Average Path Length, Connectivity and Routing Potential}
Without the involvement of GS, the six sub-constellations operate independently and remain disconnected from one another. However, once GSs are introduced, they act as bridges through FL, enabling full interconnectivity among these sub-constellations. As illustrated in Figure 6, this transforms them into a fully connected component.

This implies that, in the future—with the full deployment of ISLs—routing designers and network operators will no longer need to concern themselves with the existence of a reachable path between any two nodes. Instead, their focus can shift entirely to Quality of Service (QoS) considerations. In fact, this level of connectivity is remarkably robust, remaining largely intact even when up to 90\% of the GSs are out of service—a topic that will be explored in detail in Section~\ref{sec:Robustness}.

\begin{figure}[ht]
\centering
\includegraphics[width=0.45\textwidth ]{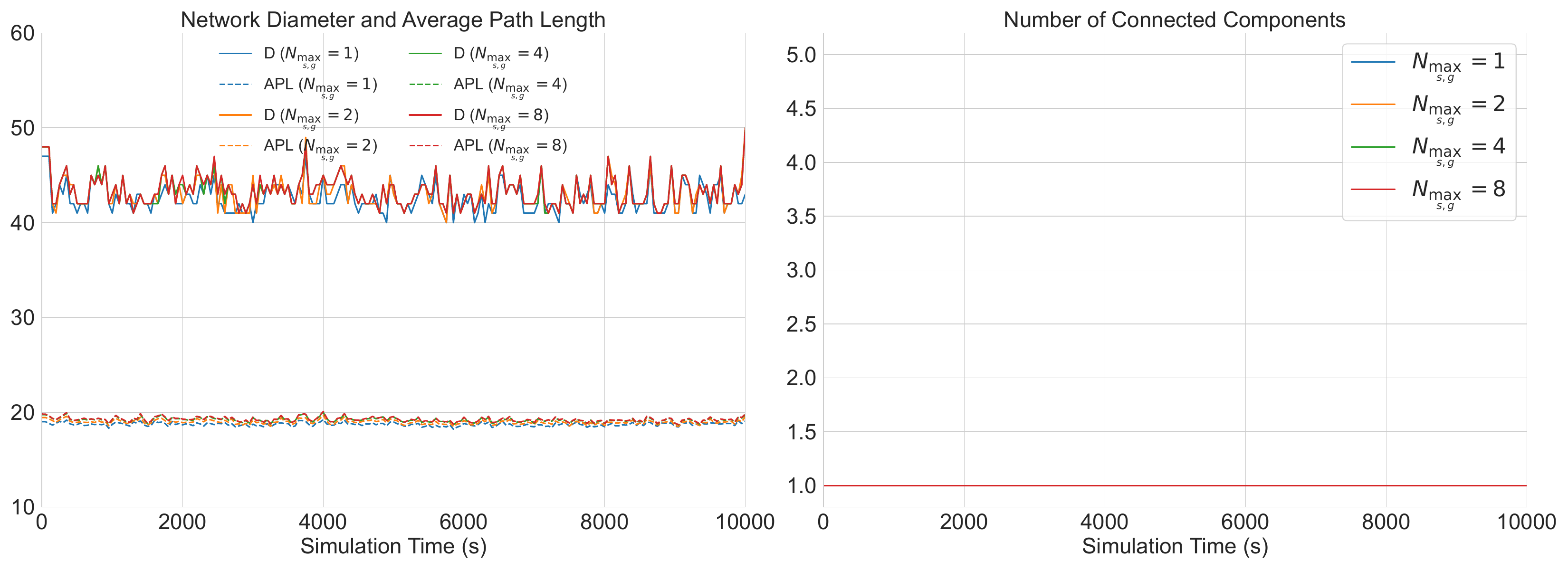}
\caption{mismatch between geographical proximity and topological distance.}
\label{fig:apl}
\end{figure}

Excitingly, with the bridging capability provided by GSs, the average path length (APL) of this ultra-large-scale network falls below 20—comparable to that of the sparsest sub-constellation, S4. The network diameter is approximately 2.2 times the APL, demonstrating \textbf{a hallmark characteristic of small-world networks} \cite{Apl_d}.

\begin{figure}[ht]
\centering
\includegraphics[width=0.45\textwidth ]{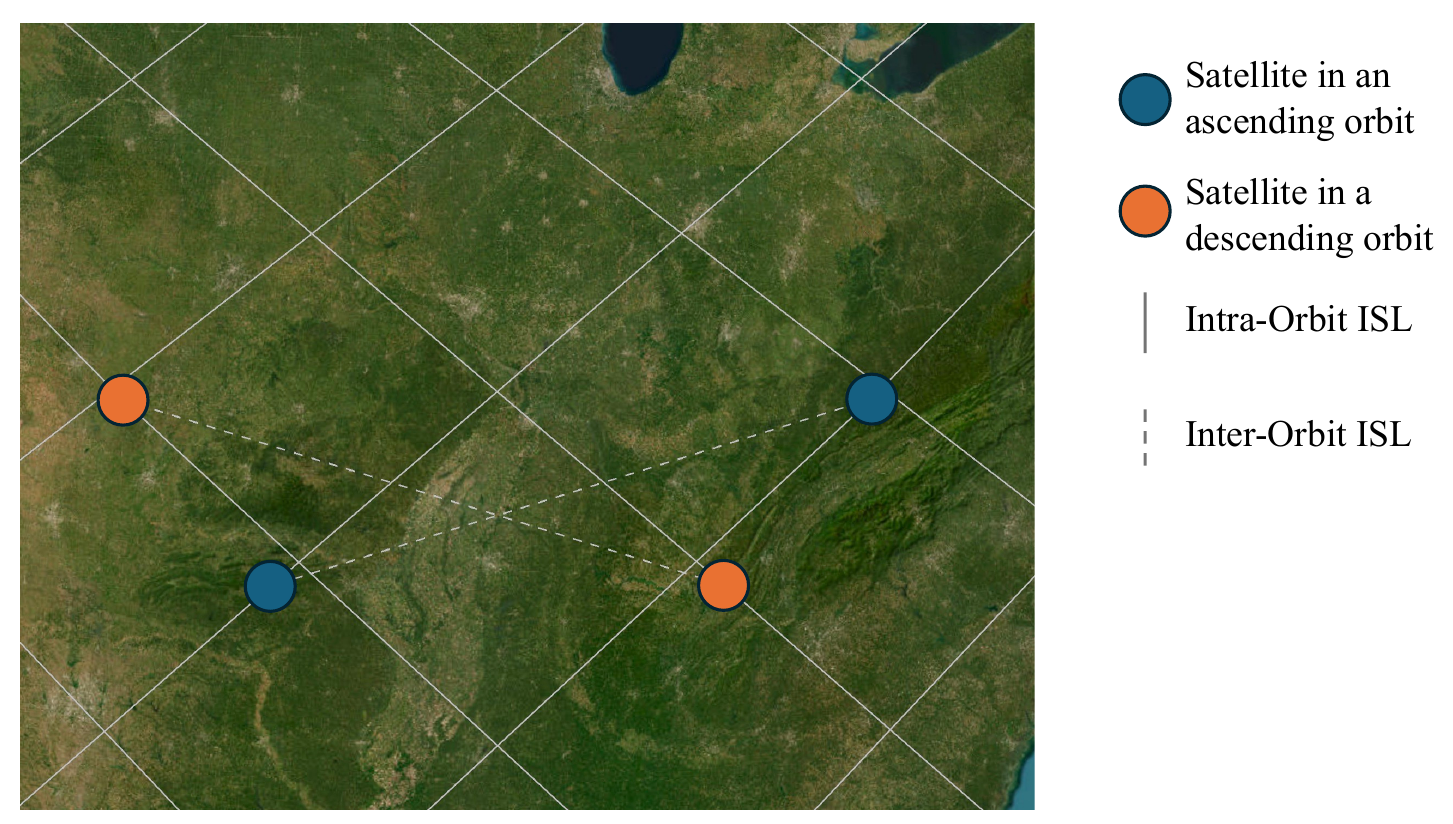}
\caption{Topological separation between geographically adjacent satellites.}
\label{fig:detour}
\end{figure}

\textbf{Explanation:} The reduction in APL is not only due to the bridging capability of GSs across different shells, but more importantly, it addresses a fundamental issue in Walker-$\delta$ constellations and mainstream ISL configuration: \textbf{the mismatch between geographical proximity and topological distance}. In each sub-constellation, the right ascensions of the ascending nodes (RAAN) of orbital planes are evenly distributed over $0^\circ$–$360^\circ$, and the arguments of perigee of satellites within each orbit are also uniformly distributed over the same range. Typically, ISLs are established only between adjacent satellites within the same orbit and between neighboring orbital planes. Near the equator, when two orbital planes have RAANs differing by $180^\circ$ and their satellites have arguments of perigee also offset by $180^\circ$, the satellites will be geographically closest yet topologically furthest apart—precisely forming the network's diameter. For a more intuitive visualization of this phenomenon, readers are referred to the animation provided in~\cite{bskempton}.

As shown in Figure~\ref{fig:detour}, satellites on ascending and descending passes—though geographically close in terms of latitude and longitude—are often topologically far apart due to the lack of direct connectivity. This is primarily because establishing stable links between such satellites is challenging: their relative motion is too rapid for reliable communication, and even if a connection is briefly established, it would be quickly lost as the satellites move out of each other’s view. However, such satellite pairs are often simultaneously within the line-of-sight of the same GS. Once both satellites establish FLs with the GS, the long-range detour required in the satellite-only topology can be effectively bypassed. This mechanism is illustrated in Figure~\ref{fig4}, where Sat1 and Sat4 communicate through a shared GS, enabling a significantly shorter path. As a result, the overall APL is significantly reduced.

Notably, the capacity of a satellite’s FL has minimal impact on either the network diameter or the APL.  This is because the GSs themselves function as the primary bridges within the network, while the satellites connected to them merely act as facilitators of this bridging role.

At present, most routing research in LEO constellation networks focuses on single-shell architectures. This approach offers both advantages and limitations. For short-distance routing—such as within a specific country or region—the single-shell model remains effective, though the ascending and descending orbit states of the entry and exit satellites must be taken into account. However, for long-haul routing, particularly intercontinental communication, leveraging GS relays becomes essential to achieve routing efficiency. This paradigm shift introduces new challenges and opportunities that warrant further investigation by the research community.

\subsection{Betweenness and Load Balancing Analysis}
Through GS relaying, both network connectivity and routing efficiency are significantly enhanced. However, SpaceX’s current GS deployment strategy is evidently driven by service availability \cite{starlink_availability_map}: GSs are predominantly established in regions with active service coverage, while areas outside these zones tend to lack GS infrastructure. This raises an important question: Could such an uneven spatial distribution of GSs result in load imbalances across the network? This subsection investigates this issue in detail.

To quantify the network-level impact, we adopt the normalized betweenness centrality \( BC(v) \) \cite{freeman1977set} of a node \( v \), defined as:
\begin{equation}
\begin{aligned}
BC_v = \frac{1}{(n-1)(n-2)} \sum_{s \neq v \neq t} \frac{\sigma_{st}(v)}{\sigma_{st}},
\end{aligned}
\label{eqBC}
\end{equation}
where:
\begin{itemize}
    \item \( n \) is the total number of nodes in the network,
    \item \( \sigma_{st} \) denotes the total number of shortest paths from node \( s \) to node \( t \),
    \item \( \sigma_{st}(v) \) denotes the number of those paths that pass through node \( v \).
\end{itemize}
 
 To assess the overall distribution pattern of betweenness centrality across the network—rather than focusing on individual GS—we introduce a divergence metric \cite{ELB} for further normalization:

\begin{equation}
\begin{aligned}
Div_{BC}= \frac{\left(\sum_{i=1}^{n_{g}} BC_{gs_{i}} \right)^{2}}{n_g \times \sum_{i=1}^{n_{g}} \left( BC_{gs_{i}} \right)^{2}},
\end{aligned}
\label{eq7}
\end{equation}
where \( n_g \) is the total number of GSs, and \( BC_{gs_i} \) is the betweenness centrality of the \( i \)-th GS. The closer the value of \( Div_{BC} \) is to 1 indicates a more uniform distribution of centrality among GSs, which reflects a more balanced network topology.

\begin{figure}[ht]
\centering
\includegraphics[width=0.45\textwidth ]{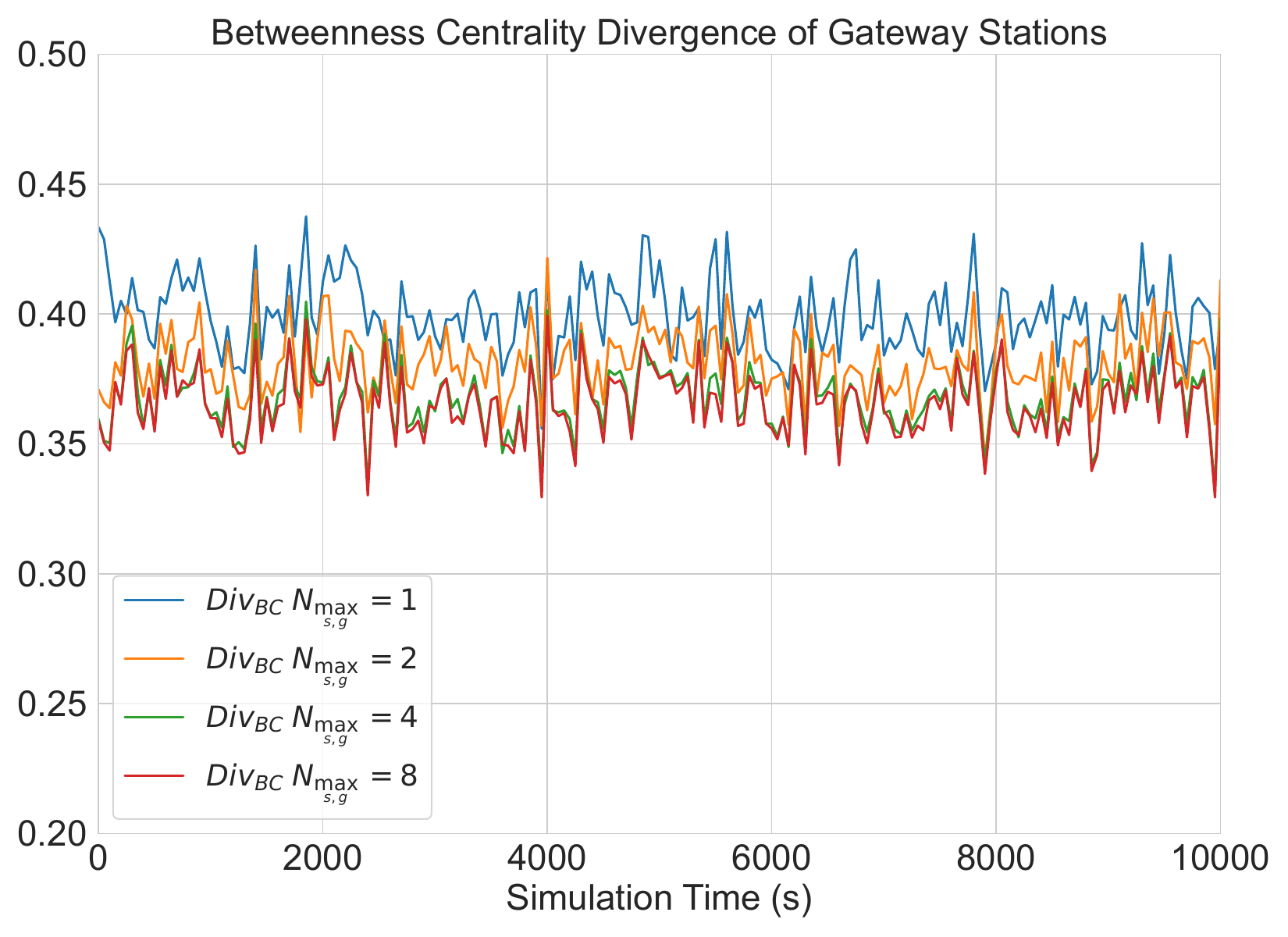}
\caption{Divergence of betweenness centrality among gateway stations.}
\label{fig:div}
\end{figure}

As Figure~\ref{fig:div} shows, there exists a significant disparity in the betweenness centrality among GSs, resulting in a consistently low $Div_{BC}$ value of approximately 0.37. This phenomenon is, in fact, quite intuitive. SpaceX’s current GS deployment is designed to be highly redundant, with extensive backup infrastructure \cite{maral2011satellite}. 

As a result, when computing shortest paths—based on propagation distance—the set of GSs involved in the unique shortest paths remains nearly constant over short time windows. GSs located near these critical relay nodes, although physically close, are not selected for routing. This leads to a concentration of centrality in a few GSs, thereby reducing the overall $Div_{BC}$ and reflecting an imbalanced load distribution.

However, this is not necessarily a negative outcome. On the contrary, it suggests that the network possesses a large number of alternative routing paths. It is foreseeable that, under a reliable and efficient GS scheduling strategy, the issue of load imbalance could be significantly alleviated. 

Although this paper does not aim to address the scheduling problem directly, incorporating load balancing mechanisms such as round-robin \cite{kleinrock1976queueing} or hash-based \cite{wang2011consistent} strategies may offer promising solutions. This observation opens up a promising direction for future work on dynamic GS load balancing and path diversification.

\subsection{Topology-Invariant Behavior in Evolving LEO Constellations}
\begin{figure*}[ht]
\centering
\includegraphics[width=0.92\textwidth ]{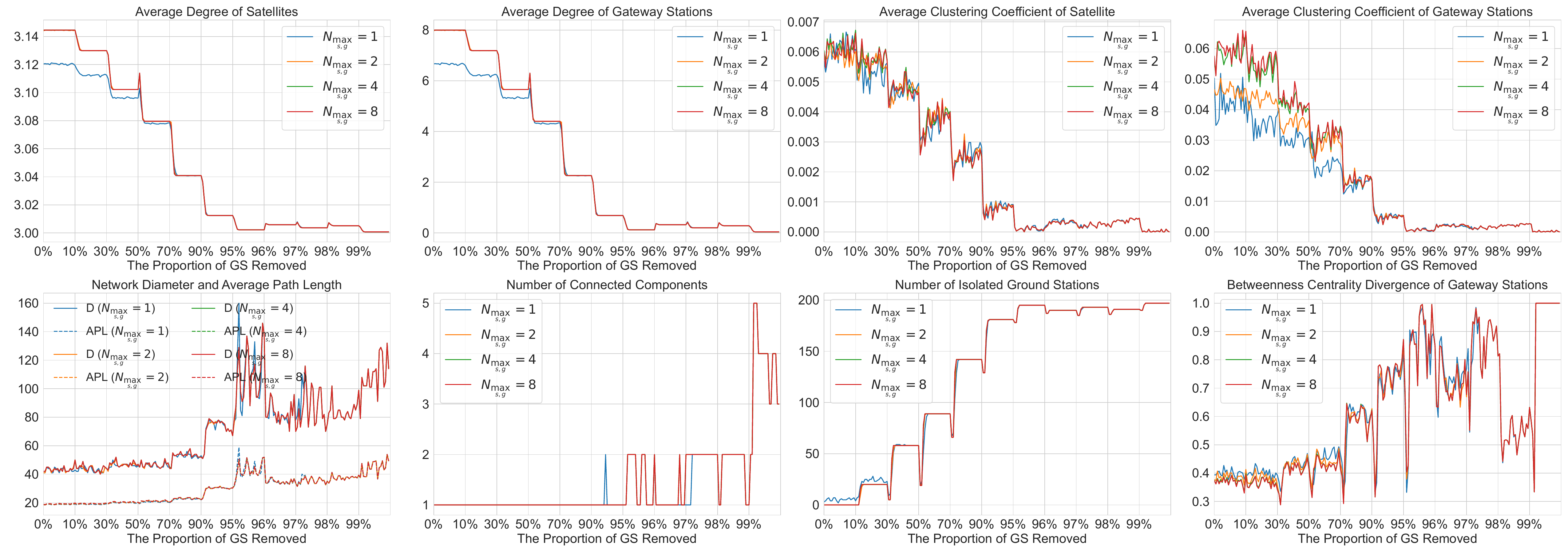}
\caption{Graph Metrics Under Progressive Gateway Station Removal.}
\label{fig:robust}
\end{figure*}
A natural question arises: do different ISL topologies or GS handover strategies affect these conclusions?

Our investigation indicates that the answer is \textbf{no}.

Current LEO constellations predominantly adopt ISL configurations based on either grid structures or the 3-way topology tested by Starlink in 2024. These configurations modify the average satellite node degree (by approximately $\pm1$), but they exert negligible influence on other network characteristics or global behavior.

Regarding GS handover strategies, we evaluated not only the default \textit{maximum elevation angle} policy but also a \textit{random selection strategy}, where the GS randomly selects a satellite whose elevation angle exceeds a given threshold ($40^\circ$) once handover is triggered. The results of graph feature remained essentially unchanged.

In fact, as the network evolves, each FL naturally exhibits varying elevation angles. Once the system reaches a steady state (approximately after 500 seconds), the network topology at any subsequent time is statistically equivalent to one resulting from randomized handovers. This confirms that our core observations are not sensitive to specific handover mechanisms or ISL layout choices.

In a similar vein, the number of shells (singe or multiple) in the constellation only slightly affects the network's APL, network diameter, and the number of isolated GSs—the latter resulting from an insufficient number of satellites and FLs capacity limitations that prevent partial GSs from establishing connections. However, the small-world property of the network remains intact.

\subsection{Rich Satellite Resource Pool Enabled by Multi-Shell Constellation Coverage}
\begin{figure}[ht]
\centering
\includegraphics[width=0.4\textwidth ]{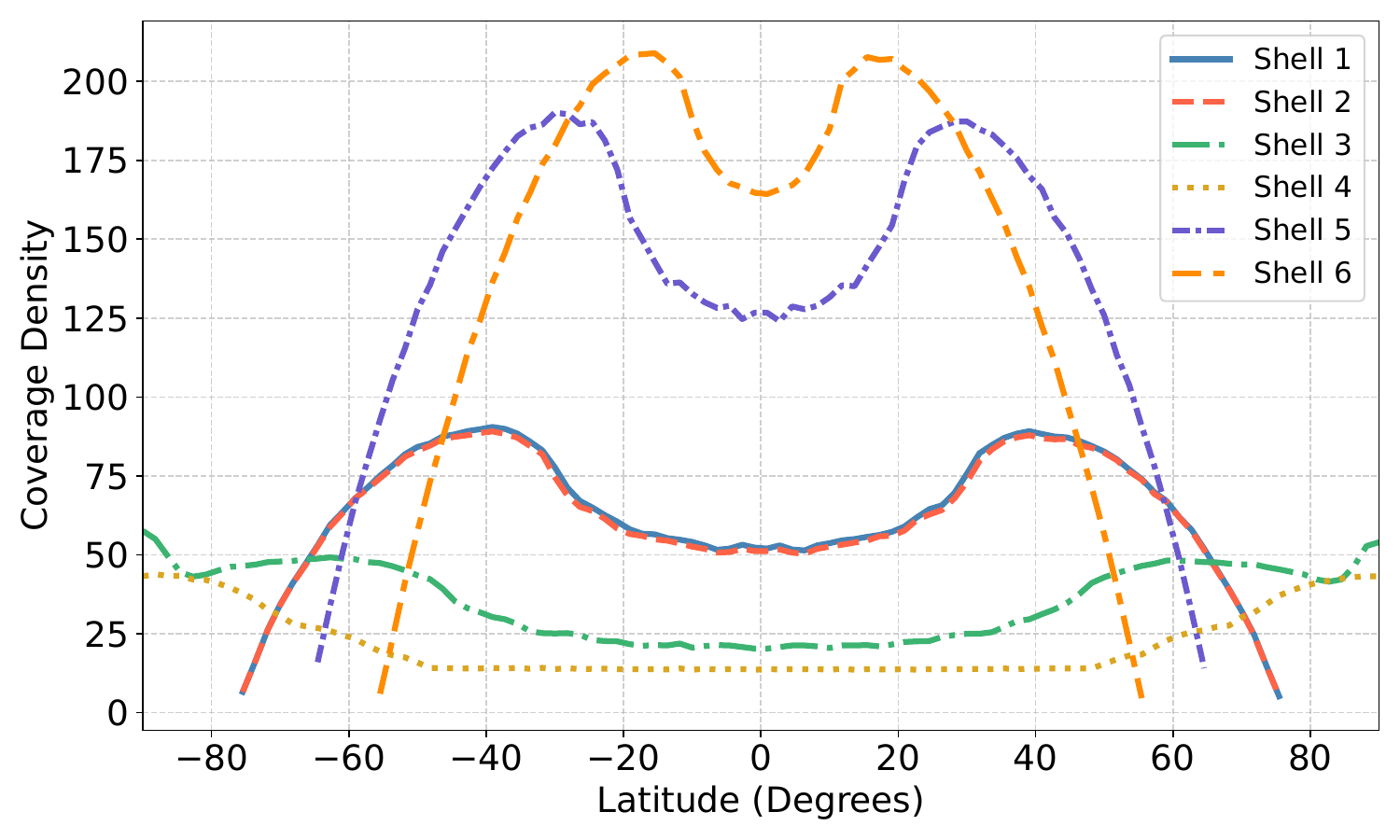}
\caption{Average coverage density vs latitude for different shells of starLink.}
\label{fig:cover}
\end{figure}

From the UT perspective, the deployment of a multi-shell LEO constellation undoubtedly enhances both service quality and overall network resilience. As illustrated in Figure~\ref{fig:cover}, the six orbital shells collectively ensure comprehensive and redundant coverage across all latitudes, including the typically underserved polar regions. 

In mid- and low-latitude regions—where the global population is most concentrated—hundreds of satellites may be simultaneously visible from a single UT at any given time. This high density of satellite visibility presents a unique combination of opportunities and challenges.

On the one hand, the abundance of accessible satellites lays a solid foundation for achieving truly global, low-latency broadband coverage. It also aligns with the long-term ambitions of operators like SpaceX, who envision satellite constellations as integral to future worldwide connectivity infrastructure. High satellite availability opens up possibilities for advanced load balancing, real-time failover, and QoS optimization.

On the other hand, the dramatic increase in satellite scale and orbital diversity significantly raises the complexity of managing satellite access, handover coordination, and mobility patterns. Traditional satellite selection and beam steering mechanisms may no longer suffice under such dense and dynamic environments. Therefore, there is a pressing need to develop more intelligent, scalable, and context-aware satellite scheduling and UT association strategies that can adapt in real-time to changing network conditions, user mobility, and varying service demands.

\section{Robustness Assessment}
\label{sec:Robustness}
The above experimental results are derived from dynamic topologies shaped by satellite mobility and GS handovers. These measurements reveal a key insight: the joint use of FLs and ISLs transforms LEO satellite constellations into small-world networks, bringing the vision of seamless global connectivity within practical reach.

This section further reveals a promising experimental finding: \textbf{the small-world nature of the network exhibits remarkable stability}. Even with the removal of 90\% of gateway stations, the small-world property persists.

The evaluation methodology is as follows: the simulation duration is set to 11,000 seconds, and every 1,000 seconds, a new set of GSs is selected randomly to access the network. The GS dropout probabilities are set to 0\%, 10\%, 30\%, 50\%, 70\%, 90\%, 95\%, 96\%, 97\%, 98\%, and 99\%, respectively. The metrics defined in Section~\ref{sec:Metrics} are measured throughout the simulation. The results are presented in Fig.~\ref{fig:robust}.

As the removal ratio increases, the number of isolated GSs rises accordingly. Meanwhile, the average degree and clustering coefficient of satellites gradually decline toward 3 and 0, respectively. Likewise, both the average degree and clustering coefficient of GSs approach zero.

When calculating \textit{Betweenness Centrality Divergence}, the denominator \( n_g \)  is set to the number of operational GSs. The resulting trend shows a steady increase, supporting our hypothesis that \textbf{SpaceX’s GS deployment features substantial redundancy}. The presence of numerous backup GSs enhances path diversity, and as the number of GSs decreases, the remaining stations increasingly function as critical network hubs.

The network’s diameter and APL remain stable when the GS removal ratio is below 90\%. However, once the removal ratio surpasses this threshold, although the ratio of network diameter to APL remains relatively constant, the overall path cost begins to rise gradually.

When the GS removal ratio reaches 90\%, concerning signs of network unreachability begin to emerge. The number of connected components can increase to two, indicating a partition in the network. Upon further investigation, this is attributed to the disconnection of the S4 shell from the rest of the sub-constellations. The main gateway stations responsible for bridging the high-inclination sub-constellation S4 are located in Alaska and Norway.

Overall, the GS removal process can be interpreted as simulating temporary outages caused by adverse weather, hardware or software failures, cyberattacks, or other unexpected incidents. Thanks to the redundancy embedded in the system design, the LEO satellite constellation network maintains strong connectivity and routing performance even when only 20 GSs remain functional.

As the number of GSs is severely reduced, a few GSs emerging as critical hubs that sustain a large portion of the interconnectivity.

\section{Discussion}
Despite the robust small-world characteristics observed in our results, several simplifying assumptions in the simulation setup may affect the extent to which these insights generalize to real-world deployments:

\textbf{FL-Based GS Access Without Demand Awareness.}
In our model, ground stations connect to satellites via feeder links based on the maximum elevation angle—a common and physically grounded heuristic. However, this method abstracts away practical considerations such as heterogeneous traffic demand, service-level agreements, and geopolitical deployment preferences. While this simplification enables a uniform analysis across global regions, it may overlook region-specific scheduling or prioritization policies. Future work could explore incorporating demand-aware or policy-driven access models to better approximate operational scenarios.

\textbf{Topology Snapshot Analysis Without Real-Time Traffic Dynamics.}
Our analysis is based on periodic topology snapshots that reveal structural characteristics such as connectivity and small-world behavior. However, we do not model packet-level dynamics, which means transient effects such as queueing, congestion propagation, or load-aware path adaptation are not captured. While such abstractions are common in structural studies, a more comprehensive understanding of end-to-end performance would benefit from integrating traffic workloads and dynamic routing protocols.

\section{Conclusion}
This paper investigates the structural and performance characteristics of large-scale LEO satellite constellations using complex network theory. Analyzing a six-shell constellation with over 10,000 satellites and nearly 200 GSs, we present several key findings with practical relevance.

First, we confirm that such constellations exhibit small-world properties—relatively high clustering and short path lengths—enabling efficient global routing. Moreover, these small-world characteristics are \textbf{intrinsic to the constellation structure}, rather than artificially engineered. Second, feeder links to GSs effectively bridge different shells, forming a unified and connected network. Third, GS relays significantly enhance routing efficiency even in large-diameter networks. Robustness analysis further shows that the network retains strong performance despite substantial GS failures. However, challenges remain. Betweenness centrality reveals imbalanced GS traffic loads, indicating a need for dynamic, traffic-aware relay strategies. These insights offer valuable guidance for designing scalable, resilient, and efficient satellite communication systems for the future.

\bibliographystyle{IEEEtran}
\bibliography{reference}

% Generated by IEEEtran.bst, version: 1.14 (2015/08/26)
\begin{thebibliography}{10}
\providecommand{\url}[1]{#1}
\csname url@samestyle\endcsname
\providecommand{\newblock}{\relax}
\providecommand{\bibinfo}[2]{#2}
\providecommand{\BIBentrySTDinterwordspacing}{\spaceskip=0pt\relax}
\providecommand{\BIBentryALTinterwordstretchfactor}{4}
\providecommand{\BIBentryALTinterwordspacing}{\spaceskip=\fontdimen2\font plus
\BIBentryALTinterwordstretchfactor\fontdimen3\font minus
  \fontdimen4\font\relax}
\providecommand{\BIBforeignlanguage}[2]{{%
\expandafter\ifx\csname l@#1\endcsname\relax
\typeout{** WARNING: IEEEtran.bst: No hyphenation pattern has been}%
\typeout{** loaded for the language `#1'. Using the pattern for}%
\typeout{** the default language instead.}%
\else
\language=\csname l@#1\endcsname
\fi
#2}}
\providecommand{\BIBdecl}{\relax}
\BIBdecl

\bibitem{1}
E.~Yaacoub and M.-S. Alouini, ``A key 6g challenge and opportunity—connecting
  the base of the pyramid: A survey on rural connectivity,'' \emph{Proceedings
  of the IEEE}, vol. 108, no.~4, pp. 533--582, 2020.

\bibitem{2}
M.~Giordani and M.~Zorzi, ``Non-terrestrial networks in the 6g era: Challenges
  and opportunities,'' \emph{IEEE Network}, vol.~35, no.~2, pp. 244--251, 2021.

\bibitem{3}
I.~{del Portillo}, B.~G. Cameron, and E.~F. Crawley, ``A technical comparison
  of three low earth orbit satellite constellation systems to provide global
  broadband,'' \emph{Acta Astronautica}, vol. 159, pp. 123--135, 2019.

\bibitem{4}
J.~C. McDowell, ``The low earth orbit satellite population and impacts of the
  spacex starlink constellation,'' \emph{The Astrophysical Journal Letters},
  vol. 892, no.~2, p. L36, apr 2020.

\bibitem{project_kuiper}
\BIBentryALTinterwordspacing
A.~Staff and T.~Kohnstamm, ``Everything you need to know about project kuiper,
  amazon’s satellite broadband network,'' Jun. 2025, last updated June 3,
  2025. [Online]. Available:
  \url{https://www.aboutamazon.com/what-we-do/devices-services/project-kuiper}
\BIBentrySTDinterwordspacing

\bibitem{10}
N.~Mohan, A.~E. Ferguson, H.~Cech, R.~Bose, P.~R. Renatin, M.~K. Marina, and
  J.~Ott, ``A multifaceted look at starlink performance,'' ser. WWW '24, New
  York, NY, USA, 2024, p. 2723–2734.

\bibitem{betz2021starlink}
\BIBentryALTinterwordspacing
E.~Betz, ``How do spacex's starlink satellites actually work?'' \emph{Discover
  Magazine}, 2021. [Online]. Available:
  \url{https://www.discovermagazine.com/technology/how-do-spacexs-starlink-satellites-actually-work}
\BIBentrySTDinterwordspacing

\bibitem{starlink2024}
\BIBentryALTinterwordspacing
SpaceX, ``Starlink progress report (v11),'' April 2024. [Online]. Available:
  \url{https://starlink-stories.cdn.prismic.io/starlink-stories/Z3QOWJbqstJ986KD_StarlinkProgress-V11_Low-Res-compressed.pdf}
\BIBentrySTDinterwordspacing

\bibitem{grid}
D.~Bhattacherjee and A.~Singla, ``Network topology design at 27,000 km/hour,''
  \emph{Proceedings of the 15th International Conference on Emerging Networking
  Experiments And Technologies}, 2019.

\bibitem{Starlink_Scheduling}
\BIBentryALTinterwordspacing
H.~B. Tanveer, M.~Puchol, R.~Singh, A.~Bianchi, and R.~Nithyanand, ``Making
  sense of constellations: Methodologies for understanding starlink's
  scheduling algorithms,'' ser. CoNEXT 2023.\hskip 1em plus 0.5em minus
  0.4em\relax New York, NY, USA: Association for Computing Machinery, 2023, p.
  37–43. [Online]. Available: \url{https://doi.org/10.1145/3624354.3630586}
\BIBentrySTDinterwordspacing

\bibitem{7}
H.~Kaushal, V.~Jain, and S.~Kar, \emph{\BIBforeignlanguage{en}{Free {Space}
  {Optical} {Communication}}}, ser. Optical {Networks}.\hskip 1em plus 0.5em
  minus 0.4em\relax New Delhi: Springer India, 2017.

\bibitem{8}
R.~Fields, C.~Lunde, R.~Wong, J.~Wicker, D.~A. Kozlowski, J.~Jordan, B.~Hansen,
  G.~Muehlnikel, W.~A. Scheel, U.~Sterr, R.~Kahle, and R.~Meyer,
  ``Nfire-to-terrasar-x laser communication results: satellite pointing,
  disturbances, and other attributes consistent with successful performance,''
  in \emph{Defense + Commercial Sensing}, 2009.

\bibitem{SpaceX2018FCC}
{SpaceX}, ``{SpaceX Non-Geostationary Satellite System, Attachment A, Technical
  Information to Supplement Schedule S},''
  \url{https://licensing.fcc.gov/myibfs/download.do?attachment_key=1569860},
  2018, accessed: Oct. 18, 2020.

\bibitem{starlinkinfo2025}
\BIBentryALTinterwordspacing
S.~Linkson, ``Starlink ground stations: What they are and how they work,''
  Starlink Info, Mar. 2025, updated March 21, 2025. About 150 ground stations
  worldwide. [Online]. Available:
  \url{https://www.starlinkinfo.com/starlink-ground-stations}
\BIBentrySTDinterwordspacing

\bibitem{starlinkinfooffical2025}
\BIBentryALTinterwordspacing
``Starlink ground station: Backbone of satellite internet,'' 2024. [Online].
  Available:
  \url{https://starlinkinstallationpros.com/starlink-ground-station-backbone-of-satellite-internet/}
\BIBentrySTDinterwordspacing

\bibitem{25}
S.~Cakaj., \emph{Ground station design and analysis for LEO satellites :
  analytical, experimental and simulation approach}.\hskip 1em plus 0.5em minus
  0.4em\relax Wiley-IEEE Press,, 2023.

\bibitem{34}
O.~Kodheli, E.~Lagunas, N.~Maturo, S.~K. Sharma, B.~Shankar, J.~F.~M. Montoya,
  J.~C.~M. Duncan, D.~Spano, S.~Chatzinotas, S.~Kisseleff, J.~Querol, L.~Lei,
  T.~X. Vu, and G.~Goussetis, ``Satellite communications in the new space era:
  A survey and future challenges,'' \emph{IEEE Communications Surveys \&
  Tutorials}, vol.~23, no.~1, pp. 70--109, 2021.

\bibitem{19}
\BIBentryALTinterwordspacing
W.~M. Wiltshire and P.~Caritj. Application for approval for orbital deployment
  and operating authority for the spacex ngso satellite system. November 15,
  2016. [Online]. Available:
  \url{https://cdn.arstechnica.net/wp-content/uploads/2016/11/spacex-Legal-Narrative.pdf}
\BIBentrySTDinterwordspacing

\bibitem{20}
M.~Handley, ``Delay is not an option: Low latency routing in space,'' in
  \emph{Proceedings of the 17th ACM Workshop on Hot Topics in Networks}, ser.
  HotNets '18, New York, NY, USA, 2018, p. 85–91.

\bibitem{22}
C.~Zhu, Y.~Li, M.~Zhang, Q.~Wang, and W.~Zhou, ``An optimization method for the
  gateway station deployment in leo satellite systems,'' in \emph{2020 IEEE
  91st Vehicular Technology Conference (VTC2020-Spring)}, 2020, pp. 1--7.

\bibitem{21}
S.~Liu, T.~Wu, Y.~Hu, Y.~Xiao, D.~Wang, and L.~Liu, ``Throughput evaluation and
  ground station planning for leo satellite constellation networks,'' in
  \emph{Space Information Networks}, Q.~Yu, Ed.\hskip 1em plus 0.5em minus
  0.4em\relax Singapore: Springer Singapore, 2020, pp. 3--15.

\bibitem{24}
M.~Kralfallah, F.~Wu, A.~Tahir, A.~Oubara, and X.~Sui, ``Optimizing the
  deployment of ground tracking stations for low earth orbit satellite
  constellations based on evolutionary algorithms,'' \emph{Remote Sensing},
  vol.~16, no.~5, 2024.

\bibitem{18}
Y.~Hauri, D.~Bhattacherjee, M.~Grossmann, and A.~Singla, ``"internet from
  space" without inter-satellite links,'' in \emph{Proceedings of the 19th ACM
  Workshop on Hot Topics in Networks}, ser. HotNets '20, New York, NY, USA,
  2020, p. 205–211.

\bibitem{28}
X.~Feng, Y.~Sun, and M.~Peng, ``Distributed satellite-terrestrial cooperative
  routing strategy based on minimum hop-count analysis in mega leo satellite
  constellation,'' \emph{IEEE Transactions on Mobile Computing}, pp. 1--16,
  2024.

\bibitem{watts1998collective}
D.~J. Watts and S.~H. Strogatz, ``Collective dynamics of 'small-world'
  networks,'' \emph{Nature}, vol. 393, no. 6684, pp. 440--442, 1998.

\bibitem{26}
R.~Albert, H.~Jeong, and A.-L. Barab{\'a}si, ``Error and attack tolerance of
  complex networks,'' \emph{Nature}, vol. 406, no. 6794, pp. 378--382, 2000.

\bibitem{27}
G.~Paul, S.~Sreenivasan, S.~Havlin, and H.~{Eugene Stanley}, ``Optimization of
  network robustness to random breakdowns,'' \emph{Physica A: Statistical
  Mechanics and its Applications}, vol. 370, no.~2, pp. 854--862, 2006.

\bibitem{satellitetoday_starlink_gen2_2022}
\BIBentryALTinterwordspacing
S.~Today, ``Fcc approves spacex's starlink gen 2 application up to 7,500
  satellites,'' Dec. 2022, accessed: 2025-06-18. [Online]. Available:
  \url{https://www.satellitetoday.com/government-military/2022/12/02/fcc-approves-spacexs-starlink-gen-2-application-up-to-7500-satellites/}
\BIBentrySTDinterwordspacing

\bibitem{starlink_demisability_2025}
\BIBentryALTinterwordspacing
SpaceX, ``Starlink approach to satellite demisability,'' 2025, accessed:
  2025-06-18. [Online]. Available:
  \url{https://www.starlink.com/public-files/Starlink_Approach_to_Satellite_Demisability.pdf}
\BIBentrySTDinterwordspacing

\bibitem{spacex_gen2_modification_2024}
\BIBentryALTinterwordspacing
L.~Space Exploration~Holdings, ``Modification of authorization for the spacex
  gen2 ngso satellite system,'' Oct. 2024, accessed: 2025-06-18. [Online].
  Available:
  \url{https://cdn.arstechnica.net/wp-content/uploads/2024/10/spacex-gigabit-application-narrative.pdf}
\BIBentrySTDinterwordspacing

\bibitem{speidel2024navigating}
\BIBentryALTinterwordspacing
U.~Speidel, ``Navigating starlink's fcc paper trail,'' Jun. 2024, accessed:
  2025-06-18. [Online]. Available:
  \url{https://blog.apnic.net/2024/06/26/navigating-starlinks-fcc-paper-trail/}
\BIBentrySTDinterwordspacing

\bibitem{array}
\BIBentryALTinterwordspacing
I.~Merino-Fernandez, S.~L. Khemchandani, J.~del Pino, and J.~Saiz-Perez,
  ``Phased array antenna analysis workflow applied to gateways for leo
  satellite communications,'' \emph{Sensors}, vol.~22, no.~23, 2022. [Online].
  Available: \url{https://www.mdpi.com/1424-8220/22/23/9406}
\BIBentrySTDinterwordspacing

\bibitem{reddit_gateway_2021}
\BIBentryALTinterwordspacing
{feral\_engineer}, ``How many gateways service one satellite?''
  \url{https://www.reddit.com/r/StarlinkEngineering/comments/qhxigc/how_many_gateways_service_one_satellite/},
  2021, accessed: 2025-06-18. [Online]. Available:
  \url{https://www.reddit.com/r/StarlinkEngineering/comments/qhxigc/how_many_gateways_service_one_satellite/}
\BIBentrySTDinterwordspacing

\bibitem{starlinkinsider2025}
\BIBentryALTinterwordspacing
S.~Insider, ``Starlink gateway locations – an updated list (2025),'' 2025,
  accessed: 2025-06-19. [Online]. Available:
  \url{https://starlinkinsider.com/starlink-gateway-locations/}
\BIBentrySTDinterwordspacing

\bibitem{29}
D.~A. Vallado, \emph{Fundamentals of Astrodynamics and Applications},
  4th~ed.\hskip 1em plus 0.5em minus 0.4em\relax Microcosm Press, 2013.

\bibitem{ns3_manual}
\BIBentryALTinterwordspacing
{The ns-3 Consortium}, \emph{{ns-3} Network Simulator}, nsnam.org, 2025.
  [Online]. Available: \url{https://www.nsnam.org}
\BIBentrySTDinterwordspacing

\bibitem{CsardiNepusz2006}
\BIBentryALTinterwordspacing
G.~Cs{\'a}rdi and T.~Nepusz, ``The igraph software package for complex network
  research,'' \emph{InterJournal, Complex Systems}, vol. 1695, 2006. [Online].
  Available: \url{https://igraph.org}
\BIBentrySTDinterwordspacing

\bibitem{36}
L.~M. Freeman, ``A set of measures of centrality based upon betweenness,''
  1977.

\bibitem{35}
M.~E.~J. Newman, ``The structure and function of complex networks,'' \emph{SIAM
  Review}, vol.~45, no.~2, pp. 167--256, 2003.

\bibitem{37}
B.~Bollob{\'a}s, ``Modern graph theory,'' in \emph{Graduate Texts in
  Mathematics}, 2002.

\bibitem{38}
J.~Liang, A.~U. Chaudhry, E.~Erdogan, and H.~Yanikomeroglu, ``Link budget
  analysis for free-space optical satellite networks,'' in \emph{2022 IEEE 23rd
  International Symposium on a World of Wireless, Mobile and Multimedia
  Networks (WoWMoM)}, 2022, pp. 471--476.

\bibitem{barabasi1999emergence}
A.-L. Barabási and R.~Albert, ``Emergence of scaling in random networks,''
  \emph{Science}, vol. 286, no. 5439, pp. 509--512, 1999.

\bibitem{erdos1960evolution}
P.~Erdős and A.~Rényi, ``On the evolution of random graphs,'' \emph{Publ.
  Math. Inst. Hung. Acad. Sci}, vol.~5, pp. 17--60, 1960.

\bibitem{Apl_d}
M.~E.~J. Newman, ``The structure and function of complex networks,'' \emph{SIAM
  Review}, vol.~45, no.~2, pp. 167--256, 2003.

\bibitem{bskempton}
\BIBentryALTinterwordspacing
B.~S. Kempton, ``A simulation tool to study routing in large broadband
  satellite networks,'' 2020. [Online]. Available:
  \url{https://search.proquest.com/docview/2438605919?accountid=27868}
\BIBentrySTDinterwordspacing

\bibitem{starlink_availability_map}
S.~. Starlink, ``Starlink availability map,''
  \url{https://www.starlink.com/map}, SpaceX / Starlink, 2025, accessed:
  2025-06-22.

\bibitem{freeman1977set}
L.~C. Freeman, ``A set of measures of centrality based on betweenness,''
  \emph{Sociometry}, vol.~40, no.~1, pp. 35--41, 1977.

\bibitem{ELB}
T.~Taleb, D.~Mashimo, A.~Jamalipour, N.~Kato, and Y.~Nemoto, ``Explicit load
  balancing technique for ngeo satellite ip networks with on-board processing
  capabilities,'' \emph{IEEE/ACM Transactions on Networking}, vol.~17, no.~1,
  pp. 281--293, 2009.

\bibitem{maral2011satellite}
G.~Maral and M.~Bousquet, \emph{Satellite communications systems: systems,
  techniques and technology}.\hskip 1em plus 0.5em minus 0.4em\relax John Wiley
  \& Sons, 2011.

\bibitem{kleinrock1976queueing}
L.~Kleinrock, \emph{Queueing Systems Volume 2: Computer Applications}.\hskip
  1em plus 0.5em minus 0.4em\relax Wiley-Interscience, 1976.

\bibitem{wang2011consistent}
Q.~Wang, M.~Liu, Z.~Zhang, Y.~Li, and X.~Zhao, ``Consistent hashing and random
  trees: Distributed caching protocols for relieving hot spots on the world
  wide web,'' \emph{IEEE INFOCOM}, pp. 1--9, 2011.

\end{thebibliography}
\end{sloppypar}
\end{document}